\documentclass{article}
\usepackage{longtable}
\usepackage{epsfig}
\usepackage{url}
\usepackage[papersize={8.5in,11in}]
{geometry}
\title{Pioneer 10/11 Telemetry Explanatory Almanac
}
\author{Viktor T. Toth\footnote{\protect\url{https://www.vttoth.com/}}}
\date{October 1, 2006}

\begin{document}

\maketitle

\begin{abstract}
This document is\footnote{Now preserved for historical purposes -- VTT, June 2026} intended to accompany our first release of edited Pioneer 10/11 telemetry data\footnote{Some of the data discussed here have been released by the Goddard Space Flight Center:\\
${}\qquad$\url{https://spdf.gsfc.nasa.gov/pub/data/pioneer/pioneer10/radio/Turyshev20170327_Pioneer-10/}\\
${}\qquad$\url{https://spdf.gsfc.nasa.gov/pub/data/pioneer/pioneer11/radio/Turyshev20170327_Pioneer-11/}}. The purpose of this data release is to provide a data set for the evaluation of the effect of on-board systematic forces on Pioneer orbits. Telemetry readings were selected for inclusion with this purpose in mind. The files presented here are all in standard formats (tab-delimited ASCII text or Microsoft Excel spreadsheets) and can be utilized using standard data editing tools.

We describe the history and availability of the raw telemetry data, documentation that can be used to interpret the data, and the tools we built in order to extract the data. We also describe script programs that were used to generate data files for release, and the manual editing process that was used to bring these data files to their final form.
\end{abstract}

\tableofcontents

\section{Introduction}

Pioneer 10 and 11, launched in 1972 and 1973, respectively, were the first spacecraft to leave the inner solar system and explore the regions beyond Mars, crossing the asteroid belt and eventually reaching Jupiter and (in the case of Pioneer 11) Saturn.

Originally designed for an approximately 2-year mission duration, the twin Pioneers far exceeded expectations. Pioneer 11 remained functional until 1995, when it was no longer possible to point the spacecraft's high-gain antenna (HGA) towards the Earth. Pioneer 10 was still sending data on the 30th anniversary of its launch, in 2002. During a contact attempt in 2003, a weak carrier signal was visible, but no telemetry was received; the last attempt to contact Pioneer 10, in 2006, was unsuccessful, as the spacecraft's onboard power supply is depleted and was no longer providing sufficient power to operate the spacecraft's radio transmitter.

Throughout most of their mission, Pioneer 10 and 11 were transmitting continuously. While the missions were active, the antennae of NASA's Deep Space Network (DSN) were tracking the spacecraft for several hours each day. During the last few years of Pioneer 10's operational life, tracking of Pioneer 10 became intermittent; nevertheless, the accumulation of data continued all the way until 2002.

The raw data received from the spacecraft contains both science results and engineering telemetry. During normal operations, science results were extracted and provided to experimenters. Engineering telemetry was used to monitor and manage the spacecraft's systems; subsequently, it was considered to be of little practical use. Consequently, while science data was preserved (indeed, much of it is available through the NSSDC archives), there were no plans for long-term retention of raw records or engineering telemetry. Data retention schedules in effect towards the end of the Pioneer 10/11 missions called for a 7-year retention of raw telemetry, after which the media was supposed to have been destroyed.

Fortunately, the telemetry from Pioneer 10/11 did not meet this fate. An early effort to preserve Pioneer 10/11 telemetry resulted in copies of magnetic tapes being made to more durable magneto-optical (``floptical'') media; these ``floptical'' cartridges were not destroyed and were still available in the early 2000s.

We originally initiated an effort to preserve this telemetry record for historical reasons. The significance of Pioneer 10/11 to the history of space exploration cannot be in doubt. We believed that the loss of the raw record of the spacecrafts' missions would have been tragic, especially when one compares it against the effort needed to preserve the record. What was once a formidable amount of data, approximately 40 gigabytes in total, today easily fits on the hard disk drive of a typical laptop computer.

The effort to transcribe all available raw telemetry from ``floptical'' cartridges to a modern computer was completed in 2005 \cite{MDR2005}. Meanwhile, we also began an effort to develop software programs with which to read telemetry records. This task is not as simple as it sounds. In order to interpret the raw telemetry, one not only needs information about the formats employed by the DSN, and the binary telemetry frame format used by the Pioneer project, we also needed calibration information, necessary to translate binary readings into physically sensible analog values. Fortunately, this information was still available in old documentation. Much of that documentation was also scheduled for destruction; indeed, we were told that the only reason why most paper files survived was that there was no budget to carry out their physical destruction!

In 2005, it became evident that the telemetry record of Pioneer 10/11 has a much more important potential use above and beyond its historical importance. It was proposed that the telemetry may help determine, with far greater accuracy than previously possible, the role that on-board systems may have played in the anomalous acceleration of the two probes, a phenomenon that is now widely known as the ``Pioneer anomaly'' \cite{JPL1998,JPL2002}. Previous, qualitative arguments about the role of propellant system leaks and thermal radiation can now be made more precise, and most significantly, the telemetry may help determine the temporal evolution of these on-board effects, thus providing a means through which to correlate these effects with Doppler acceleration data.

In the following, we provide details on the Pioneer spacecraft (Section~\ref{sec:spacecraft}), their missions (Section~\ref{sec:missions}), Pioneer telemetry (Section~\ref{sec:telemetry}), the process and tools used to recover telemetry data (Section~\ref{sec:tools}), and the data products we extracted from telemetry files (Section~\ref{sec:data}).

\section{The Pioneer 10/11 spacecraft and missions}
\label{sec:spacecraft}

Pioneer 10, the first spacecraft to travel the solar system beyond the orbit of Mars, was launched on March 3, 1972 on top of an Atlas-Centaur launch vehicle. The launch vehicle placed the spacecraft on a trajectory aimed for an encounter with the planet Jupiter in December, 1973. Pioneer 10 completed this journey as scheduled, performing the first ever close-up observations of the gas giant and then continuing towards the outer solar system. In addition to numerous scientific observations, Pioneer 10 also demonstrated that it is possible for spacecraft to safely cross the asteroid belt and survive the intense radiation environment of Jupiter.

Pioneer 10's sister craft, Pioneer 11, was launched almost exactly one year later, on April 5, 1973. Its trajectory was similar to that of Pioneer 10, targeting an encounter with Jupiter in December, 1974. Encouraged by the success of Pioneer 10, a decision was made to alter the orbit of Pioneer 11 in April 1974, causing it to fly closer to Jupiter during its flyby, performing a gravity assist maneuver that placed it on a trajectory for an eventual encounter with Saturn. That encounter did, in fact, take place in September, 1979, providing the first ever close-up observations of the ringed planet.

Following planetary encounters, the twin Pioneers continued to travel towards the outer parts of the solar system.

\subsection{The Pioneer spacecraft}

From available documentation, we were able to extract relevant information about the spacecrafts' geometry and appearance, thermal control, electrical, communication, and propulsion subsystems.

\subsubsection{Physical description}

\begin{figure*}[t!]
\centering
\psfig{figure=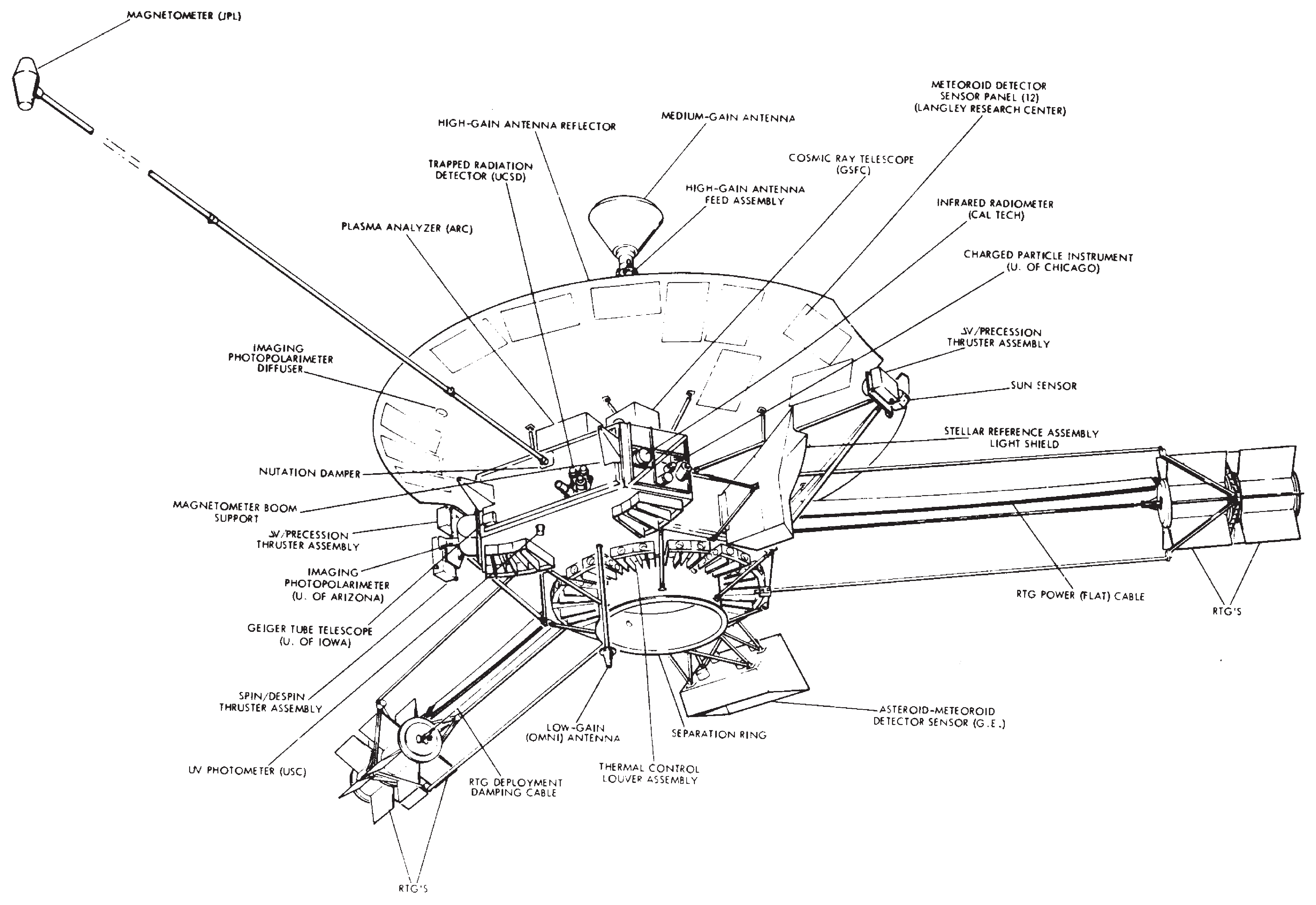,width=\linewidth}
\caption{\noindent A drawing of the Pioneer spacecraft.\label{fig:pioneer}}
\end{figure*}

The appearance of the Pioneer spacecraft (Figure~\ref{fig:pioneer}) is dominated by the 2.74~m diameter high-gain antenna. Behind the antenna is the main spacecraft body. The body consists of two compartments. The larger compartment, shaped like a regular hexagonal block, houses most of the spacecraft's electrical systems, including its power regulation and power distribution electronics and its attitude control subsystem.

At the center of this larger compartment is the spacecraft's $\sim$36~l spherical fuel tank. The fuel tank has two compartments separated by a flexible membrane: one side is filled with hydrazine monopropellant fuel, while the other side contains N$_2$ pressurant.

Adjoining to the main compartment is a smaller compartment, also shaped as an irregular hexagonal block. This compartment houses most science instruments.

Outside the two compartments are situated several science instrument components and sensors. Also external to the main compartment are the spacecraft's battery and the heat radiating plate of its shunt circuit.

A characteristic feature of the Pioneer spacecrafts' appearance is the set of three booms, positioned 120$^\circ$ apart, extending several meters from the center of the spacecraft. Two of these booms, $\sim$3~m in length, each hold a set of SNAP-19 radioisotope thermoelectric generators, the spacecrafts' primary electrical power source. The third boom, $\sim$6~m in length, holds a magnetometer.

\begin{figure*}[t!]
\centering
\psfig{figure=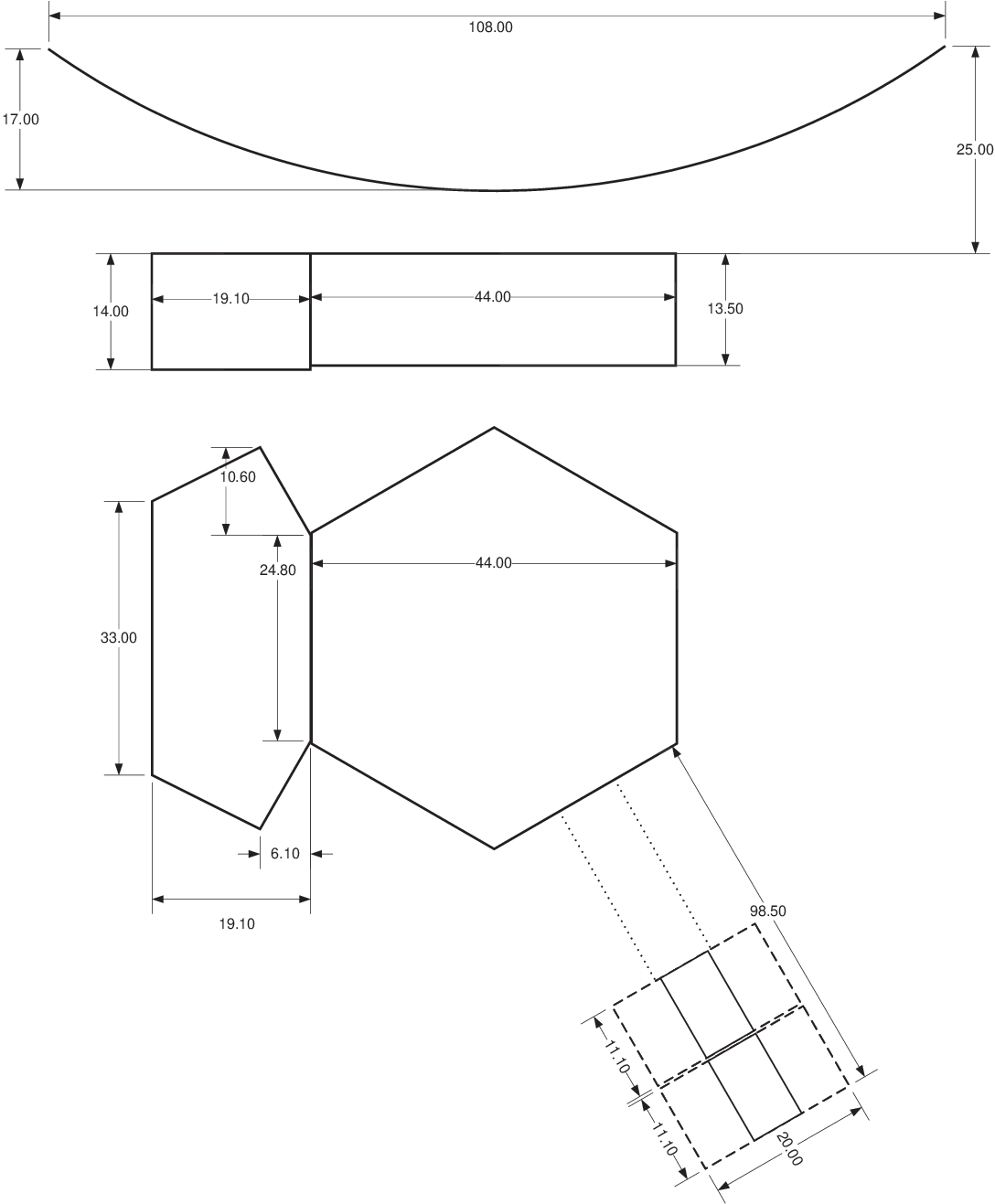,width=0.67\linewidth}
\caption{\noindent Approximate dimensions of the main components of the Pioneer spacecraft body (from \cite{PC220}). The HGA, main spacecraft body, and one RTG shown; RTG boom length not drawn to scale.\label{fig:geometry}}
\end{figure*}

Absent detailed engineering drawings of the spacecraft, we were nevertheless able to obtain approximate dimensions of the main spacecraft components. In particular, \cite{PC220} contained several partially dimensions engineering drawings that were drawn to scale, and from which other dimensions, not marked, could be measured with acceptable precision. Figure~\ref{fig:geometry} shows the geometry of the spacecraft with dimensions obtained from \cite{PC220}.

The total mass of the Pioneer 10/11 spacecraft was approximately 250~kg, of which 27~kg was propellant. (actual launch masses were 256.51 kg for Pioneer 10, 258.32 kg for Pioneer 11, including propellant and pressurant mass of 25.76 kg and 27.67~kg, respectively \cite{PFG100151}.)

\subsubsection{Thermal control}

The thermal control capabilities of the Pioneer spacecraft were designed to maintain acceptable operating temperatures inside the spacecraft compartments both in the near-Earth environment, dominated by thermal radiation from the Sun, and also in the extreme cold of interplanetary space beyond the orbit of Jupiter.

Thermal control on board Pioneer 10/11 relied almost entirely on passive systems. The main thermal control elements are
\begin{itemize}
\item Thermal control blankets covering the exterior of the spacecraft compartments;
\item A bimetallic spring activated radiative louver system;
\item Several (11) appropriately positioned small (1~W) radioisotope heating units (RHUs);
\item A small (2~W) electrical heater system for the spacecraft's battery;
\item Electrical heaters (1~W each) for the two fuel lines connecting the spacecraft's fuel tank with its thruster cluster assemblies.
\end{itemize}

Of these, only the battery heater and line heater assemblies could be commanded from the ground. The functioning of the remaining systems could only be verified indirectly, e.g., by observing the temperature readings of the affected subsystems.

Heat sources affecting the thermal behavior of the Pioneer spacecraft were either external or internal. The main external heat source is, of course, the Sun; during planetary encounters, however, thermal radiation from the planet may also have had a minor effect on the spacecraft's thermal behavior.

\begin{table}
\caption{RTG total power measurements prior to launch. The power measurement uncertainty is believed to be 0.1~W (\cite{SNAP19}).\label{tb:rtgheat}}
\vskip 6pt
\centering
\begin{tabular}{|ccccc|}
\hline\hline
RTG\#&Spacecraft&Location&Test date&Thermal power (W)\\\hline
44&Pioneer 10&Outboard&October 1971&649\\
45&Pioneer 10&Inboard&November 1971&646\\
46&Pioneer 10&Outboard&November 1971&647\\
48&Pioneer 10&Inboard&December 1971&649\\
49&Pioneer 11&Outboard&September 1972&649\\
51&Pioneer 11&Inboard&October 1972&650\\
52&Pioneer 11&Outboard&October 1972&649\\
53&Pioneer 11&Inboard&October 1972&649\\
\hline
\end{tabular}
\end{table}

Internally generated heat was nuclear, electrical, or chemical in origin. The largest heat source on board were the four RTGs, which generated approximately 2.5~kW of waste heat at the beginning of mission. However, as the RTGs were situated at the end of long booms, only a small fraction of this heat was expected to be intercepted by the spacecraft itself; most RTG heat would be radiated away into space directly. The total thermal power of the RTGs (NB: This is the sum of power radiated away in the form of waste heat and power removed in the form of electrical energy) is known \cite{SNAP19}. Table~\ref{tb:rtgheat} summarizes the thermal power of all 8 ``flight'' RTGs.

The spacecraft's maneuvering thrusters were also producing heat when they were operating. During an extended firing, a thruster could easily reach temperatures in excess of 500$^\circ$C, warming up thruster cluster assemblies and nearby spacecraft structural elements. As thruster firings were rare and most maneuvers consisted only of a few brief thruster pulses, the overall thermal effect of maneuvers on the spacecraft was likely not very significant.

\begin{figure*}[t!]
\centering
\psfig{figure=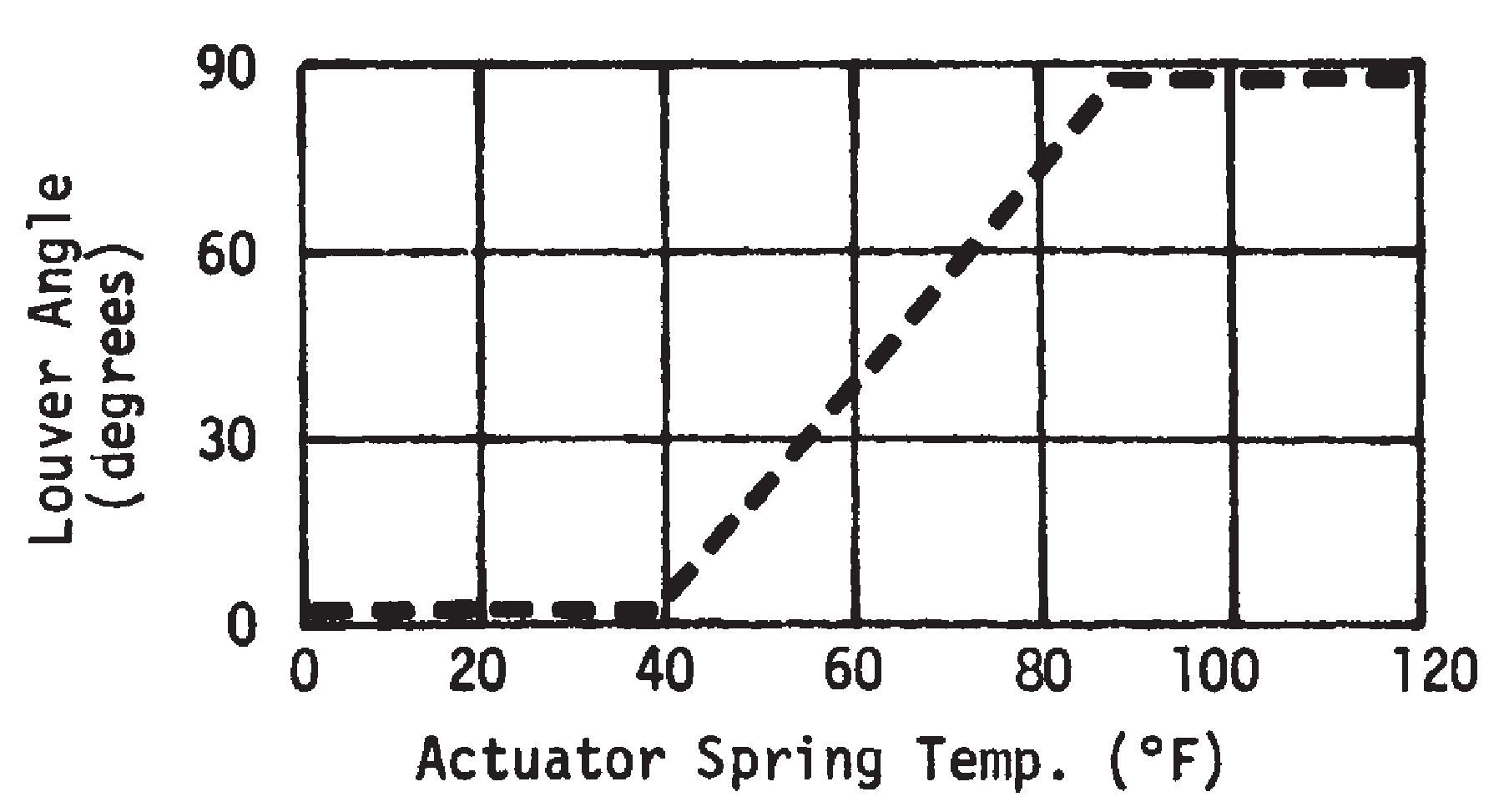,width=0.5\linewidth}
\caption{\noindent Louver angle as a function of platform temperature. (from \cite{PC202}).\label{fig:louver}}
\end{figure*}

The last main heat source on board was the collection of the spacecrafts' electrical equipment, consuming more than 100~W of electrical early in the mission, and converting most of that electrical power into heat. As most of this heat was produced inside the well insulated spacecraft body, it was necessary to provide a mechanism through which this heat could escape. Leaving the compartment open to space was, however, not an option, as this would have resulted in unacceptable cold temperatures later in the mission, when solar heating was insignificant and the amount of available electrical power would have diminished. Therefore, a louver system was used on the rear side of both hexagonal compartments of the spacecraft body. The louvers were activated by bimetallic springs that were thermally (radiatively) coupled to the electrical platform beneath. The louvers were designed to be fully open when the platform temperatures exceeded 85$^\circ$F, and were to close fully at platform temperatures below 40$^\circ$ (Figure~\ref{fig:louver}).

\begin{table}
\caption{Results of pre-launch thermal vacuum chamber testing of louver assemblies. Values are emitted thermal power, in W, per louver assembly as a function of electronics platform temperature behind the louvers. From \cite{TCSDR3}.}
\label{tb:louvers}
\begin{center}
\begin{tabular}{|c|c|c|}\hline\hline
$T_{\mathrm{platform}}\,(^\circ\mathrm{F})$&2-blade assembly&3-blade assembly\\\hline\hline
100&9.0&14.1\\
85&7.8&12.3\\
60&4.0&7.3\\
40&1.7&3.0\\
20&1.4&2.1\\\hline\hline
\end{tabular}
\end{center}
\end{table}

The amount of thermal radiation escaping through the louver system, as a function of the platform temperature, is known. A thermal vacuum chamber test was performed on the individual louver assembly elements (2-blade and 3-blade louver assemblies) and its results are still available (Table~\ref{tb:louvers}).

\begin{table}
\caption{Pioneer 10/11 surface radiometric properties: Solar absorptance ($\alpha$) and infrared emittance ($\epsilon$).}
\begin{center}
\begin{tabular}{|l|c|c|c|}\hline\hline
Surface&Area ($\mathrm{m}^2$)&$\alpha$&$\epsilon$\\\hline\hline
HGA front&5.91&0.21&0.85\\
HGA rear&5.91&0.17&0.04\\
Spacecraft body, front&1.53&0.17&0.70\\
Spacecraft body, rear&1.24&0.40&0.70\\
Spacecraft body, RTG sides&0.65&0.40&0.70\\
Spacecraft body, other sides&1.21&0.17&0.70\\
RTG fin surfaces&&0.20&0.83\\
Louver blades (closed)&0.29&0.17&0.04\\\hline\hline
\end{tabular}
\label{tb:radiometric}
\end{center}
\end{table}

Solar and infrared radiometric properties of the main Pioneer 10/11 spacecraft surfaces are summarized in Table~\ref{tb:radiometric}.

Many temperature readings from on board the spacecraft appear in the telemetry. Most of these readings are not very useful if one is trying to determine the average temperature of the spacecraft compartments or parts thereof, as they are specific readings of equipment or subsystems near heat sources. However, several sensors exist the temperatures of the electronics platforms; other sensors exist that measure the ``fin root'' temperature of each RTG, providing at least a qualitative indicator of the RTGs' thermal history.

\begin{figure}
\centering
\psfig{figure=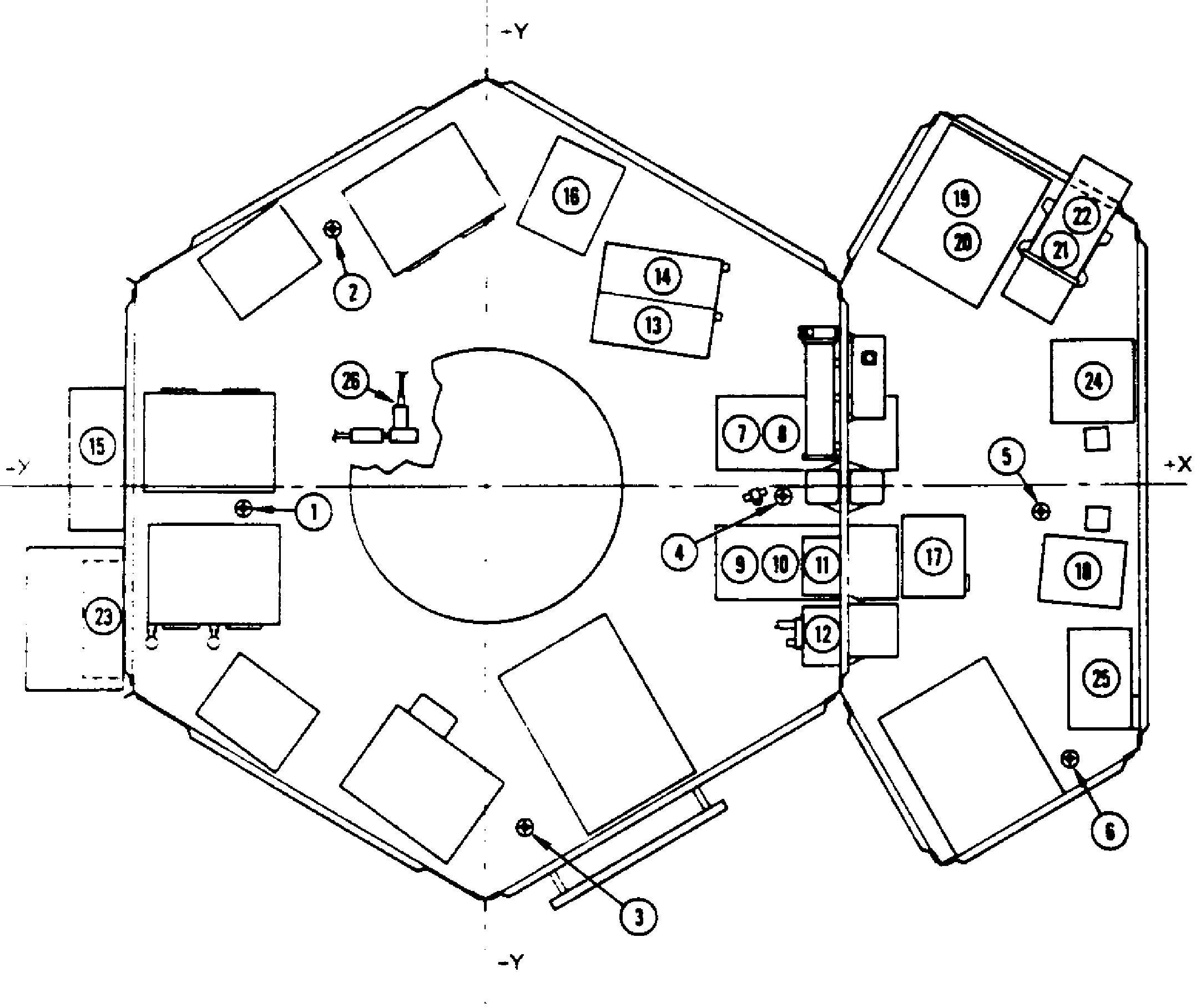,width=0.6\linewidth}
\caption{\noindent Location of thermal sensors in the instrument compartment of the Pioneer 10/11 spacecraft \cite{PC202}. Platform temperature sensors are mounted at locations 1 through 6.\label{fig:tempsens}}
\end{figure}

Figure~\ref{fig:tempsens} shows the location of many temperature sensors inside the spacecraft body.

\subsubsection{Electrical subsystem}

The primary source of electrical energy on board the Pioneer 10/11 spacecraft was a set of 4 SNAP\footnote{ Space Nuclear Ancillary Power}-19 radioisotope thermoelectric generators. The nominal power of each generator was 40~W, providing a total of 160~W of electrical power at the beginning of mission.

The RTGs were fueled with $^{238}$Pu isotope fuel. The nuclear fuel heated the hot ends of several bimetallic thermocouples; the cold ends of the thermocouples were attached to exterior radiating fins that radiated the waste heat of the RTGs away into space.

The efficiency of the thermocouples was low. The thermal power of a freshly fueled RTG was approximately 650~W, of which $\sim$610~W was radiated away in the form of waste heat.

The amount of heat available was expected to decay in accordance with the known decay rate of the $^{238}$Pu fuel, at a half life of 87.74 years. The amount of electrical power available from the RTGs was expected to decay faster, due to decreasing thermocouple efficiency as the temperature difference between the hot and cold ends decreased and also due to degradation of the thermocouples themselves.

\begin{figure*}[t!]
\centering
\psfig{figure=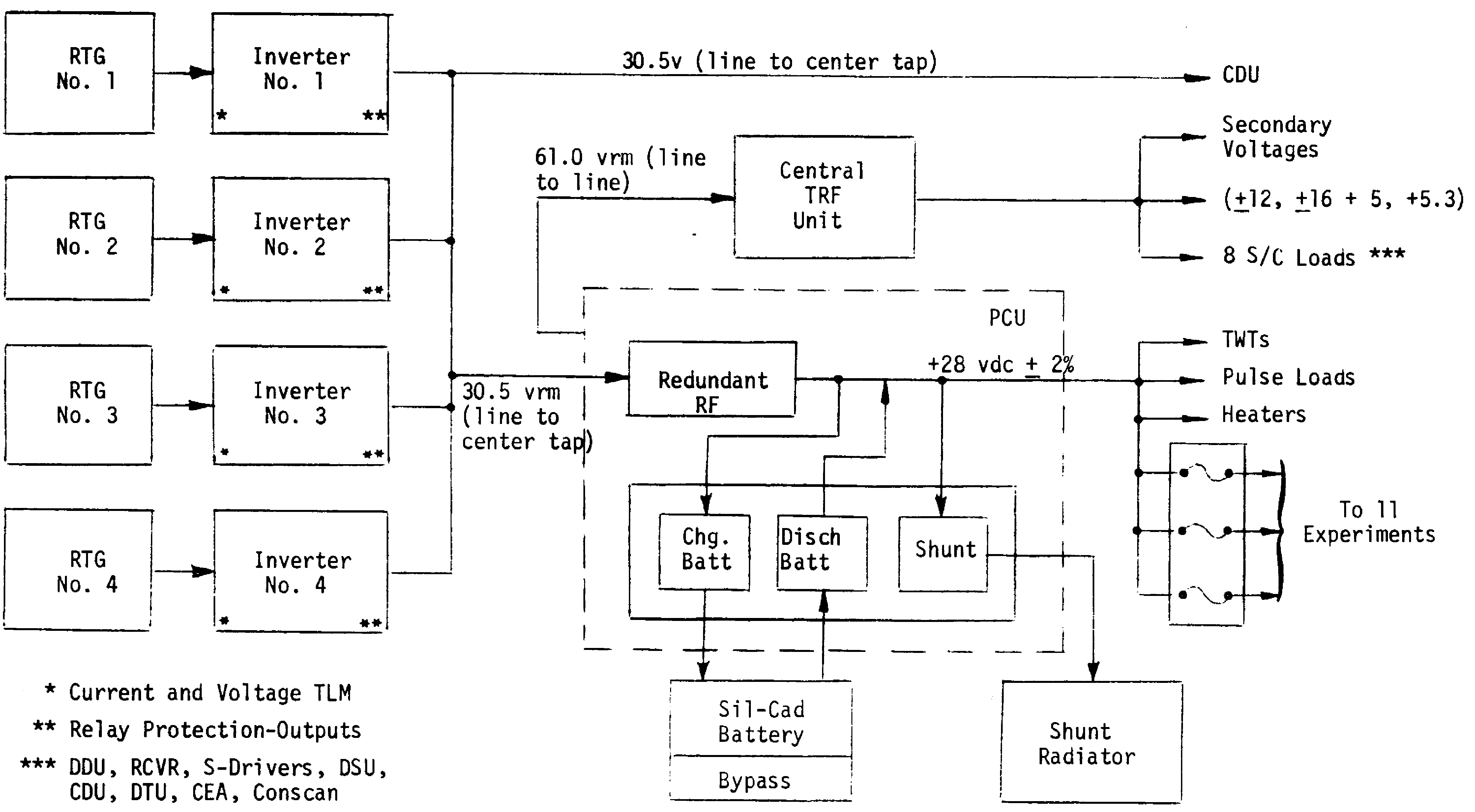,width=\linewidth}
\caption{\noindent An overview of the Pioneer 10/11 electrical subsystem (from \cite{PC202}).\label{fig:electrical}}
\end{figure*}

The output of each RTG was DC current at $\sim$4~V, $\sim$10~A at the beginning of mission. This DC current was converted to AC by four inverter circuits (one per RTG; see Figure~\ref{fig:electrical}). The output of the inverters was a 30.5~V (61~V peak-to-peak) alternating current which was fed to several power supply circuits. One power supply circuit generated the 28~V DC main bus voltage that provided primary power to most spacecraft subsystems and science instruments; another power supply circuit provided regulated low voltage outputs (e.g., +5~V, $\pm$12~V, $\pm$16~V).

The life of the thermocouples was maximized if the thermocouples always operated at optimum efficiency. This, however, results in a discrepancy between electrical power available from the RTGs vs. electrical power consumed by the spacecraft's subsystems. Excess electrical power was routed to a shunt circuit that dissipated excess power in the form of heat. Some of that heat was dissipated by the shunt regulator circuit inside the spacecraft body, while the rest of the heat was emitted by a shunt radiator plate, mounted on the outside, on one side of the spacecraft body.

The spacecraft also had a silver-cadmium battery on board. The battery was continuously monitored and charged. Battery power was available to support peak loads, when the spacecraft's electrical power consumption briefly exceeded the amount of available power from the RTGs.

\subsubsection{Communication}

The Pioneer spacecraft were designed to be spin-stabilized, with the HGA continuously pointing towards the Earth. Most communication between the Earth and the spacecraft were expected to utilize the HGA.

For operations immediately following launch in the near-Earth environment, when the HGA was not yet properly oriented, two additional antenna systems were provided: a medium gain antenna (MGA) in the form of a horn mounted at the focal point of the HGA, and an omnidirectional low-gain antenna (LGA) mounted at the rear of the spacecraft.

The spacecraft had two transmitters and two receivers.

The transmitters were of the travelling wave tube (TWT) type, transmitting at a microwave frequency of $\sim$2292~MHz. Switching from one transmitter to the other was accomplished via ground command. A microwave relay routed the output of one TWT to the HGA, and the other to the MGA and LGA (connected together.) The state of this relay was controlled by ground command; individual TWT amplifiers were also powered up or down by ground command.

The spacecrafts' receivers operated at a frequency of $\sim$2111~MHz. Another microwave relay was used to determine which receiver was connected to the HGA, and which was connected to the MGA and LGA (connected together.) The microwave relay could be operated by ground command, and also by an on-board watchdog circuit that, unless inhibited by ground command, caused a switch from one receiver to another after an extended period of receiving no signal from the Earth.

The transmitter had two modes of operation. It could either utilize an on-board crystal oscillator to synthesize the transmitter frequency, or it could operate in so-called ``coherent mode''. The ratio of the transmitted vs. the received frequency was 240/221. In coherent mode, the transmitter frequency was synthesized from the frequency of the received carrier at this exact rational ratio. In this mode of operation, the accuracy and stability of the transmitted frequency was not limited by the accuracy and stability of any oscillator on board, and therefore, coherent mode was usable for precision Doppler observations.

Late in its mission, the transmitter on board Pioneer 10 could no longer operate using its on-board oscillator, only in coherent mode. Therefore, an uplinked signal was necessary in order to receive a downlink.

The spacecrafts' communication subsystem also provided an attitude control mechanism called ``CONSCAN'' (conical scan). A mechanical system consisting of a piston that was operated by electrically heated freon gas could displace the microwave feed from the focal point of the HGA. If the HGA was not pointed exactly at the Earth, this introduced a slight amplitude modulation into the received signal at the frequency of the spacecraft's spin rate. A digital signal processor on board translated this modulation into firing signals for the spacecrafts' velocity and precession control thrusters. By appropriately timed signals, the spacecraft could reorient itself, ``homing in'' on the carrier signal received from the Earth.

\subsubsection{Thrusters and propulsion}

\begin{figure*}[t!]
\centering
\psfig{figure=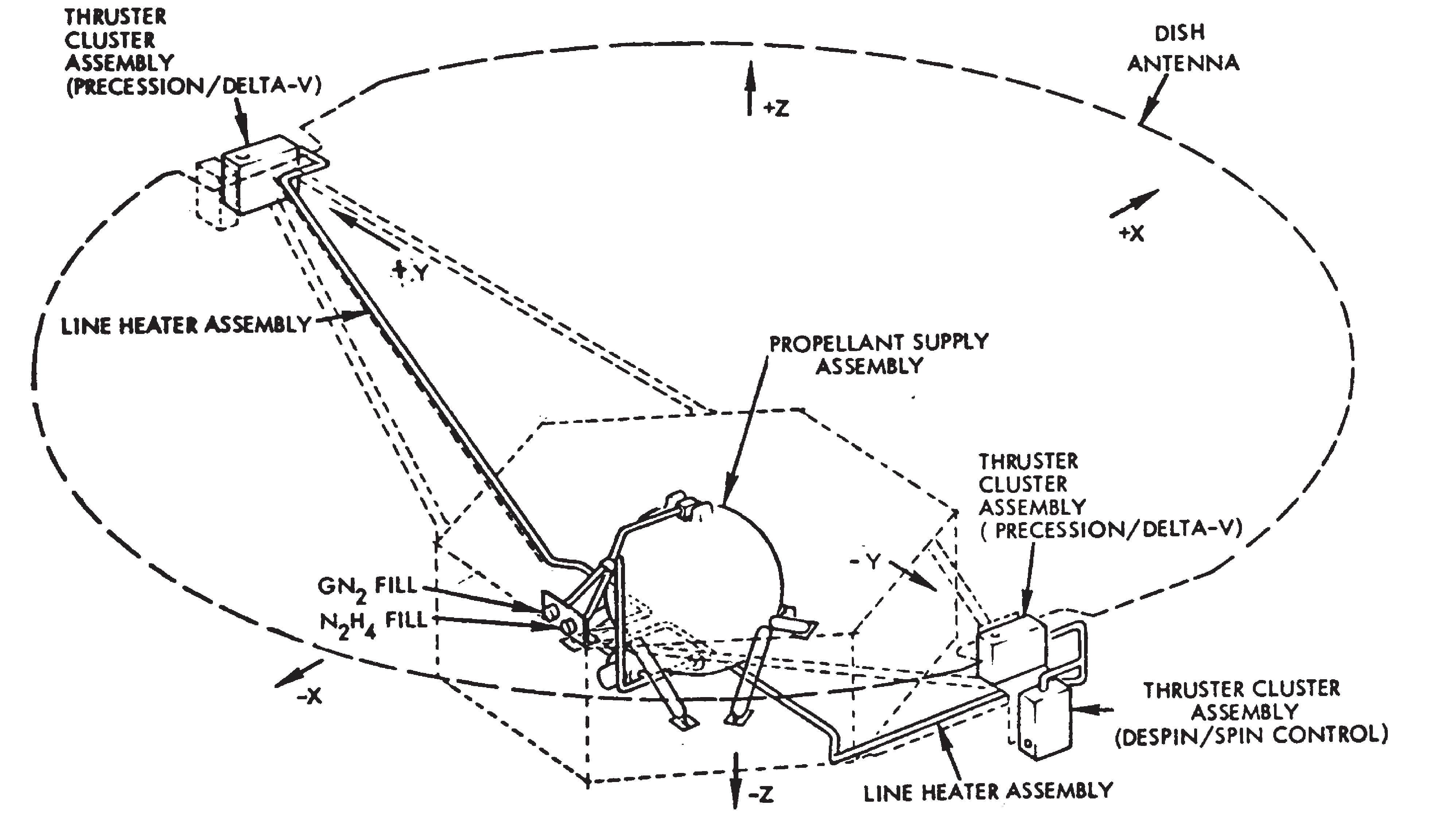,width=0.8\linewidth}
\caption{\noindent An overview of the Pioneer 10/11 propulsion subsystem (from \cite{PC202}).\label{fig:propulsion}}
\end{figure*}

The Pioneer 10/11 spacecraft each had 6 thrusters housed in three thruster cluster assemblies (Figure~\ref{fig:propulsion}). All thruster cluster assemblies were mounted near the rim of the HGA. Two of the thrusters were used for spin control, the remaining four for attitude and precession control.

The Pioneer 10/11 spacecraft were designed to perform three types of maneuvers:
\begin{itemize}
\item{\it Spin/despin maneuvers} were used, at the beginning of mission, to alter the spacecraft's spin rate. One or the other thruster of the spin/despin thruster cluster assembly was used to effect a spin-up or spin-down of the spacecraft.
\item{\it Delta-v ($\mathrm{\Delta v}$) maneuvers} were used to change the spacecraft's trajectory. With the spacecraft pointing in the right direction, two velocity and precession thrusters (VPTs) on opposite sides of the high gain antenna (HGA), pointing in the same direction, were fired, accelerating the spacecraft along the spin axis.
\item{\it Precession maneuvers} were used to change the orientation of the spacecraft's spin axis. Two velocity and precession thrusters, mounted on opposite sides of the HGA and firing in opposing directions at a specific roll phase of the spacecraft could be used to precisely alter the orientation of the spacecraft spin axis.
\end{itemize}

The two spin control thrusters were aimed along the circumference of the antenna rim, in opposite directions. Firing one thruster increased the spacecraft's spin; firing the other decreased the spin. Spin control thrusters were used at the beginning of the mission to adjust the spin rate of the spacecraft to approximately the nominal spin rate of 4.8~RPM. This spin rate was achived on Pioneer 10; on Pioneer 11, however, a thruster malfunction caused an anomalous spin rate increase, and the spacecraft's spin remained near or over 8~RPM throughout its mission.

The velocity and precession thrusters (VPTs) were also mounted near the rim of the HGA, on two sides that were 180$^\circ$ apart. The two thrusters in each cluster were oriented perpendicular to the antenna plane, in opposite directions. VPTs were always fired in pairs, one from each thruster cluster. When two VPTs pointing in the same direction were fired, the spacecraft's velocity changed. When two VPTs pointing in opposite directions were fired, this caused the spacecraft's spin axis to precess.

Thruster firings could take place under ground control, or as part of the closed-loop ``CONSCAN'' maneuver that provided a means for the spacecraft to automatically ``home in'' on the DSN carrier signal received from the Earth. In all cases, every time a thruster was fired, the corresponding thruster pulse count register was incremented by one.

VPTs were numbered 1 through 4, or alternatively, labeled as follows:

\begin{center}
\begin{tabular}{c@{: }c}
VPT 1&1A\\
VPT 2&1B\\
VPT 3&2B\\
VPT 4&2A\\
\end{tabular}
\end{center}

VPTs were used in the following combinations:

\begin{center}
\begin{tabular}{r@{: }c}
``Positive'' precession&2 and 3 (1B and 2B)\\
``Negative'' precession&1 and 4 (1A and 2A)\\
Fore $\mathrm{\Delta}$v&1 and 3 (1A and 2B)\\
Aft $\mathrm{\Delta}$v&2 and 4 (1B and 2A)\\
\end{tabular}
\end{center}

\begin{table}
\caption{Pioneer 10/11 propulsion system capabilities.\label{tb:propulsion}}
\vskip 6pt
\centering
\begin{tabular}{|lcrr|}\hline\hline
Maneuver&Thruster mode&Maximum&Propellant\\\hline
Despin&Steady state&59~rpm&0.64~lb\\
Precession&Pulse&1250$^\circ$&8.08~lb\\
$\Delta v$&Steady state&200~m/s&49.08~lb\\
Spin control&Pulse&14~rpm&1.10~lb\\
Angular calibration&Pulse&90$^\circ$&0.60~lb\\
Unavailable&\multicolumn{2}{c}{Trapped/leaked propellant}&1.00~lb\\\hline
TOTAL&&&60.50~lb\\\hline\hline
\end{tabular}
\end{table}

The capabilities of the Pioneer 10/11 propulsion system are summarized in table~\ref{tb:propulsion}.

\subsection{The Pioneer 10/11 missions}
\label{sec:missions}

The primary mission objective for Pioneer 10/11 was to cross the asteroid belt and reach the planet Jupiter.

\subsubsection{Trajectory}

Pioneer 10 was launched on March 3, 1972 on top of an Atlas-Centaur launch vehicle. The powered flight phase of the mission lasted 13 minutes and 44 seconds, placing the spacecraft on a trajectory that carried it to an encounter with Jupiter on December 3, 1973 (closest approach.) Subsequently, Pioneer 10 was leaving the solar system in a direction opposite to the motion of the Sun through the interstellar medium.

Pioneer 11 was launched on April 5, 1973, also on top of an Atlas-Centaur launch vehicle. Its trajectory was to take it to Jupiter, for a closest approach on December 4, 1974. Encouraged by the success of Pioneer 10 and by the minimum amount of damage that spacecraft suffered as it flew through Jupiter's radiation belts, the trajectory of Pioneer 11 was modified on April 19, 1974. This modification enabled the spacecraft to fly closer to Jupiter's cloudtops, performing a gravity assist maneuver that placed it on a trajectory for an encounter with Saturn. That encounter took place on September 1, 1979. Subsequently, Pioneer 11 was leaving the solar system in the direction of the Sun's motion through the interstellar medium.

\begin{table}
\caption{Major milestones of the Pioneer 10/11 projects.\label{tb:milestones}}
\vskip 6pt
\centering
\begin{tabular}{|l|c|c|}
\hline\hline
Event&Pioneer 10&Pioneer 11\\\hline
Launch&March 3, 1972&March 6, 1973\\
Orbit correction&N/A&April 19, 1973\\
Jupiter encounter&December 3, 1973&December 4, 1974\\
Saturn encounter&N/A&September 1, 1979\\
Last telemetry&April 27, 2002&September 30, 1995\\
\hline\hline
\end{tabular}
\end{table}

\begin{figure*}[t!]
\centering
\psfig{figure=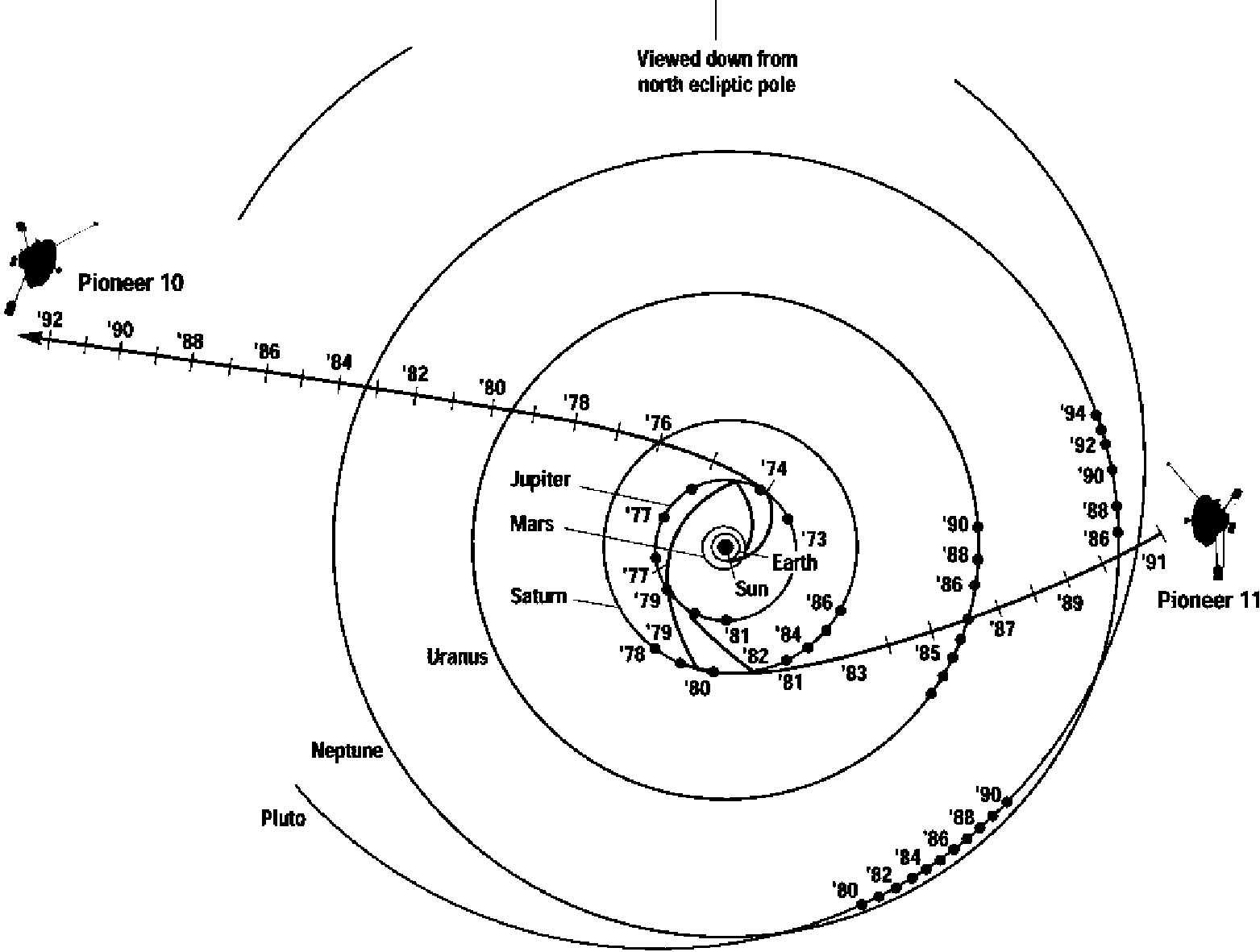,width=\linewidth}
\caption{\noindent Trajectories of Pioneer 10 and 11.\label{fig:trajectory}}
\end{figure*}

The major milestones of the Pioneer 10/11 missions are summarized in Table~\ref{tb:milestones}. Figure~\ref{fig:trajectory} shows the trajectory of both spacecraft in the solar system.

\subsubsection{Spacecraft operations}

The initial Pioneer 10/11 flight operations after launch included despin, boom deployment, and orientation. When the spacecraft were separated from the launch vehicle, they were spinning at approximately 60~rpm. The spin/despin thrusters were used to reduce this spin rate to approximately 15~rpm. At that time, the booms holding the RTGs and the magnetometer were deployed. This deployment took place in the plane containing the spacecraft's center of gravity and perpendicular to the spin axis, and therefore, it did not alter the orientation of the spacecraft, but it did cause its rate of spin to further decrease. After successful boom deployment, the spin/despin thrusters were once again used to finalize the spin rate.

The spacecraft was also oriented to ensure that its HGA pointed towards the Earth. Calibration maneuvers were also performed to precisely estimate the spacecrafts' response to solar radiation pressure and to characterize thrusters.

During the flight (cruise) phase of the mission, the spacecraft were using their instruments to measure the interplanetary medium. Occasionally, trajectory correction maneuvers were performed. A typical trajectory correction maneuver required three operations. First, the spacecraft's spin axis had to be changed to ensure that the spacecraft points in the desired direction; second, the spacecraft's velocity was increased or decreased; and third, the spin axis was reoriented to ensure that the HGA again points towards the Earth.

Throughout the missions, it was necessary to regularly adjust the spacecrafts' spin axis to ensure that the spacecrafts continue to point towards the Earth. This was accomplished in one of two ways. Either the closed-loop CONSCAN maneuver was used, which required only an initiation command and a continuously present carrier signal from the Earth; or, calculated thruster pulses were fired by ground command. Such a ground-controlled precession maneuver was preferred because it was possible to execute it with a smaller loss of propellant.

\section{Pioneer telemetry}
\label{sec:telemetry}

The Pioneer spacecraft were designed to transmit continuously, at a data rate of 16--2048 bits per second (bps), through their 2.74~m HGA. Transmissions from Pioneer 10/11 were received by the DSN. The nominal transmitter power on board the spacecraft was 8~W. The low transmitter power, combined with the extreme distances between the spacecraft and the Earth, necessitated the use of large (70~m) DSN antennae in order to communicate with the spacecraft and achieve a received signal level of $\sim -$180~dBm near the end of mission. Despite limited availability of these DSN facilities, several hours of telemetry was received from both Pioneers until very late in their missions. Coverage was effectively continuous throughout the mission of Pioneer 11 (1973--1995); for Pioneer 10, coverage was continuous until the formal end of the mission in 1998, while data was received sporadically afterwards, until the last telemetry data point in 2002.

Even at low bit rates, this represents a tremendous amount of data. The total amount of Pioneer 10/11 telemetry in uncompressed form amounts to approximately 40 gigabytes of data, or roughly 1 billion telemetry frames. It is remarkable that such a large amount of data survived this many years, despite any lack of plans for long-term telemetry data retention (indeed, plans actually called for the destruction of this data, which fortunately was not carried out.)

\subsection{Data Organization}

\begin{table}
\caption{Overview of the Master Data Record (MDR) format. See Table~\ref{tb:mdrdefs} for field definitions.}
\label{tb:mdr}
\centering
\vskip 12pt
{\tt
\begin{tabular}{r|c|c|c|c|}
Word&Byte 1&Byte 2&Byte 3&Byte 4\\\hline
1&\multicolumn{4}{|c|}{TIME TAG}\\\cline{2-5}
2&SC/ID&TIME COR FLAG&\multicolumn{2}{|c|}{DAY OF YEAR}\\\cline{2-5}
3&UDT&DDT&SYNC COND CODE&DQI\\\cline{2-5}
4&\# BIT ERRORS PN&YEAR DIGIT&\multicolumn{2}{|c|}{SNR}\\\cline{2-5}
5&DSS&LOCK STATUS BITS&\multicolumn{2}{|c|}{CONFIGURATION INDICATORS}\\\cline{2-5}
6&SPCL DATA TYPE&GDD&\multicolumn{2}{|c|}{\# OF DATA BITS IN RECORD}\\\cline{2-5}
7&\#AGC SAMP AVER&HSD ERR CON BITS&\multicolumn{2}{|c|}{RATE OF DATA TRANSMISSION}\\\cline{2-5}
8&\multicolumn{4}{|c|}{AVERAGE AGC OVER DATA IN RECORD}\\\cline{2-5}
9&FORMAT&SPARE&\multicolumn{2}{|c|}{NUMBER OF FRAMES}\\\cline{2-5}
10&\multicolumn{4}{|c|}{ZEROS}\\\cline{2-5}
11&\multicolumn{4}{|c|}{}\\
.&\multicolumn{4}{|c|}{}\\
.&\multicolumn{4}{|c|}{192 BIT FRAME ONE}\\
.&\multicolumn{4}{|c|}{}\\
16&\multicolumn{4}{|c|}{}\\\cline{2-5}
17&\multicolumn{4}{|c|}{}\\
.&\multicolumn{4}{|c|}{}\\
.&\multicolumn{4}{|c|}{192 BIT FRAME TWO}\\
.&\multicolumn{4}{|c|}{}\\
22&\multicolumn{4}{|c|}{}\\\cline{2-5}
23&\multicolumn{4}{|c|}{}\\
.&\multicolumn{4}{|c|}{}\\
.&\multicolumn{4}{|c|}{192 BIT FRAME THREE}\\
.&\multicolumn{4}{|c|}{}\\
28&\multicolumn{4}{|c|}{}\\\cline{2-5}
29&\multicolumn{4}{|c|}{}\\
.&\multicolumn{4}{|c|}{}\\
.&\multicolumn{4}{|c|}{192 BIT FRAME FOUR}\\
.&\multicolumn{4}{|c|}{}\\
34&\multicolumn{4}{|c|}{}\\\cline{2-5}
35&\multicolumn{2}{|c|}{MS CLOCK LSB'S (FRAME 2 OF 4)}&\multicolumn{2}{|c|}{MS CLOCK LSB'S (FRAME 3 OF 4)}\\\cline{2-5}
36&\multicolumn{2}{|c|}{MS CLOCK LSB'S (FRAME 4 OF 4)}&\multicolumn{2}{|c|}{DDA INFO (1)}\\\cline{2-5}
37&\multicolumn{2}{|c|}{DDA INFO (1) cont}&\multicolumn{2}{|c|}{GROUND RECEIVER AGC}\\\cline{2-5}
38&\multicolumn{4}{|c|}{DDA INFO (2)}\\\cline{2-5}
39&\multicolumn{2}{|c|}{SPARE}&\multicolumn{2}{|c|}{DDA INFO (3)}\\\cline{2-5}
40&\multicolumn{2}{|c|}{DDA INFO (3) cont}&\multicolumn{2}{|c|}{SPARE}\\\cline{2-5}
41&\multicolumn{4}{|c|}{DDA INFO (4)}\\\cline{2-5}
42&SCF 1&SCF 2&SCF 3&SCF 4\\\cline{2-5}
\end{tabular}
}
\end{table}

When the DSN facility achieved a ``lock'' on the received signal, it began to record the data received. The recorded data was packaged in the form of ``Master Data Records'' (MDRs), which contained information about the DSN facilities as well as the raw telemetry frames that were received from the spacecraft. These records were copied to magnetic tape, which were then shipped to the Pioneer project for further processing.

\subsubsection{File Organization}

Today, data that was once stored on computer tape and later transcribed to ``floptical'' cartridges is now available in the form of computer files. The data is organized such that one file per spacecraft is used for each day of operation. File names follow a naming convention that encodes the spacecraft identifier (Pioneer 10 = 23, Pioneer 11 = 24), the year, and the day of the year: for instance, {\tt m2372063.mdr} is the MDR file for Pioneer 10, for DOY 63 in 1972 (launch day.)

MDR files are stored in folders that also follow a consistent naming convention. Folder names encode the spacecraft ID and the year. For some years, multiple folder were used to keep the number of files stored in any single folder reasonably low. Multiple folder for the same spacecraft and year were numbered sequentially. Thus for instance, {\tt 23P7201} is the first folder for 1972 for Pioneer 10.

This organization makes it possible to quickly locate a file for a certain date. The file name can be synthesized from the spacecraft identifier, year, and day of year. (NB: The day of year is 1-based; DOY 1 is January 1. This is in contrast with conventions used elsewhere, notably in UNIX software libraries, where the day of the year is zero-based.) To locate the file for a specific DOY, several directories need to be searched. Starting with the first numbered directory for the year. The search stops either when the desired file is located, or when, as files are enumerated in alphabetic order, a file name is found that succeeds the desired file name, which indicates that there is no data for the desired date.

Conceivably, the search could be made more effective by a binary search algorithm, but the small number of directories means that any improvement in performance would likely be marginal at best.

\subsubsection{Master Data Records}

The MDR was a format used at the DSN for storing spacecraft telemetry, in particular telemetry from Pioneer 10/11.

The MDR consisted of a header, a telemetry data area, and a footer (Table~\ref{tb:mdr}).

The MDR header is 320 bits in length. The header is followed by a data area of 768 bits. Following the data area is the MDR footer, itself 256 bits in length.

MDR data is stored in big-endian byte format. (On Intel-architecture processors, this necessitates an appropriate swapping of bytes to ensure that data is read correctly.)

The MDR size is fixed, at $320+768+256=1344$ bits, or 168 bytes. This is regardless the number of telemetry frames that is present in an MDR. If fewer than four telemetry frames are present, the record is padded with zeroes. The number of valid telemetry frames in an MDR is stored in the MDR header.

The MDR header contains information recorded by the DSN station. This includes the ground received time of the telemetry record, information about DSN station configuration, and information about the quality of the signal received.

Many of these fields were only relevant at the time of reception. Other fields are essential for the decoding and reconstruction of telemetry data. In particular, the MDR header contains the exact timestamp when the telemetry frame(s) contained within the MDR were received. The MDR header also tells us how many telemetry frames are contained in the record and how they are organized.

\begin{small}
\begin{longtable}{|p{0.25\linewidth}p{0.65\linewidth}|}
\caption{\normalsize MDR data word definitions (from \cite{ARC221}).\label{tb:mdrdefs}}\\
\hline\hline
Defined Term&Definition\\\hline
\endfirsthead
\caption[]{(continued)}\\\hline
Defined Term&Definition\\\hline
\endhead
\hline
\endfoot
\hline\hline
\endlastfoot
{\tt TIME TAG}&Elapsed time in milliseconds since the start of the day (ground received UTC.)\\
\vskip -6pt{\tt SC/ID}&\vskip -6pt Spacecraft identifier. Value is 23$_\mathrm{dec}$ for Pioneer 10, 24$_\mathrm{dec}$ for Pioneer 11. Other values were used for simulation purposes.\\
\vskip -6pt{\tt TIME COR FLAG}&\vskip -6pt A value of FF$_\mathrm{hex}$ indicates that the time tag may be suspect or corrected.\\
{\tt DAY OF YEAR}&Number of days since the beginning of the year. January 1 = 1.\\
\vskip -6pt{\tt UDT}&\vskip -6pt User data type code.\\
\vskip -6pt{\tt DDT}&\vskip -6pt Data-dependent type code.\\
\vskip -6pt{\tt SYNC COND CODE}&\vskip -6pt Sync condition code.\\
&{\tt ~~~~0}: No sync\\
&{\tt ~~~~1}: Trailing code only\\
&{\tt ~~~10}: Leading code only\\
&{\tt ~~~11}: Full sync\\
&{\tt ~~111}: Full sync (leading code only was requested and found.)\\
\vskip -6pt{\tt DQI}&\vskip -6pt Data quality indicator.\\
&{\tt ~~~0}: No sync\\
&{\tt ~~~1}: SNR is below specified minimum, high-speed data (HSD) block contains bit errors.\\
&{\tt ~~10}: Either SNR is below minimum or HSD block contains bit errors.\\
&{\tt ~~11}: All indicators good.\\
\vskip -6pt{\tt \# BIT ERRORS PN}&\vskip -6pt Number of bit errors detected in leading identification code.\\
\vskip -6pt{\tt YEAR DIGIT}&\vskip -6pt Last two digits of the year in BCD.\\
\vskip -6pt{\tt SNR}&\vskip -6pt Signal to noise ratio in dB, encoded as a 12-bit fixed position binary value with 6 bits to the right of the binary point.\\
\vskip -6pt{\tt DSS}&\vskip -6pt DSN station identifier.\\
\vskip -6pt{\tt LOCK STATUS BITS}&\vskip -6pt Lock status bits extracted from HSD. A value of 1 in a bit position indicates no lock.\\
&~~bits 0-1: 0\\
&~~bit 2: Receiver in lock\\
&~~bit 3: Demodulator in lock\\
&~~bit 4: Bit sync in lock or not in use\\
&~~bit 5: Symbol sync in lock or not in use\\
&~~bit 6: Data decoder assembly in lock or not in use\\
&~~bit 7: Block decoder in lock.\\
\vskip -6pt{\tt CONFIGURATION INDICATORS}&\vskip -6pt DSN configuration bits extracted from HSD.\\
\vskip -6pt{\tt SPCL DATA TYPE}&\vskip -6pt Special data type code.\\
\vskip -6pt{\tt GDD}&\vskip -6pt Gross data descriptor.\\
&{\tt ~~0}: Real time transmission\\
&{\tt ~~1}: Non-telemetry data replay by DSS\\
&{\tt ~~2}: All system data replay by GCF\\
&{\tt ~~4}: Telemetry data (digital) replay by DSS\\
&{\tt ~~5}: Telemetry data (analog) replay by DSS\\
\vskip -6pt{\tt \# OF DATA BITS IN RECORD}&\vskip -6pt 192 or 384 for synced data. May be less than 192 for non-synced data.\\
\vskip -6pt{\tt \#AGC SAMP AVER}&\vskip -6pt Number of AGC samples included in the average AGC value.\\
\vskip -6pt{\tt HSD ERR CON BITS}&\vskip -6pt Error condition bits from HSD. 0 = bad, 7 = good.\\
\vskip -6pt{\tt RATE OF DATA TRANSMISSION}&\vskip -6pt Rate of data transmission.\\
\vskip -6pt{\tt AVERAGE AGC OVER DATA IN RECORD}&\vskip -6pt Average AGC over the data record. Fixed-point binary value with the binary point between bits 28 and 29.\\
\vskip -6pt{\tt FORMAT}&\\
\vskip -6pt{\tt NUMBER OF FRAMES}&\vskip -6pt Number of telemetry frames:\\
&{\tt~~0}:1 192-bit frame\\
&{\tt~~4}:2 192-bit frames\\
&{\tt~~8}:3 192-bit frames\\
&{\tt~12}:4 192-bit frames.\\
\vskip -6pt{\tt MS CLOCK LSB'S}&\vskip -6pt Least significant bits of time stamp for frames 2--4.\\
\vskip -6pt{\tt DDA INFO}&\vskip -6pt Digital data acquisition information.\\
\vskip -6pt{\tt SCF}&\vskip -6pt SCID correction flag for frames 1--4.\\
\end{longtable}
\end{small}

MDR files contain MDRs in time order. As the MDR size is fixed, this makes it possible to rapidly locate the MDR for a particular time using an efficient binary search algorithm.

\subsubsection{Telemetry Frames}

The size of each telemetry record or frame was 192 bits. In most formats, 48 bits of each frame were reserved for identification, synchronization, and subcommutator information; the remaining 144 bits were available for data transmission.

The first 24 bits of each telemetry frame contained the following data:

\paragraph{Bits 1--2: Mode ID} The data handling subsystem was designed to operate in one of three modes. In mode 00, real-time mode, data from instruments is transmitted without intermediate storage. In mode 01, telemetry store mode, consecutive frames are simultaneously transmitted and stored in the spacecraft's 49,152 bit data storage unit (DSU) until the DSU is full. In mode 10, memory readout mode, telemetry information is read back from the DSU and transmitted. (Mode 11 was not used.)
\paragraph{Bit 3: DTU select} A value of 0 in this bit position indicates that DTU A is being used.
\paragraph{Bits 4-6: Bit rate} The bit rate is $16\cdot 2^n$ where $n$ is the numeric value of the 3-bit word at this bit position. The lowest bit rate is 16~bps, the highest is 2048~bps.
\paragraph{Bits 7-24: Frame sync} The frame synchronization bits contain the bit pattern\\ 111~100~110~101~000~000.
\vskip 12pt
The first 24 bits of the second half of each telemetry frame (i.e., bits 97--120) were also reserved. They contained the following data:

\begin{table}[h]
\caption{Telemetry frame format identifiers.\label{tb:formats}}
\vskip 6pt
\centering
\begin{tabular}{|cc|}
\hline\hline
Format&Value\\\hline
{\tt A}&{\tt 0~1~0~0~X}\\
{\tt B}&{\tt 0~0~0~0~X}\\
{\tt C1}&{\tt 0~X~1~0~0}\\
{\tt C2}&{\tt 0~X~1~0~1}\\
{\tt C3}&{\tt 0~X~1~1~0}\\
{\tt C4}&{\tt 0~X~1~1~1}\\
{\tt A/D1}&{\tt 1~1~0~0~0}\\
{\tt A/D2}&{\tt 1~1~0~0~1}\\
{\tt A/D3}&{\tt 1~1~0~1~0}\\
{\tt B/D1}&{\tt 1~0~0~0~0}\\
{\tt B/D2}&{\tt 1~0~0~0~1}\\
{\tt B/D3}&{\tt 1~0~0~1~0}\\
\hline
\end{tabular}
\end{table}

\paragraph{Bits 97--101: Format ID} The format of the current frame (see Table~\ref{tb:formats}). Note that this information is also encoded in the MDR header.
\paragraph{Bits 102--108: Subcommutator ID} The subcommutator ID is a 7-bit binary value that runs from 0--127. It serves as an index to one of the 128 engineering subcommutator words in the C format (0=$C_{101}$, 127=$C_{432}$). The lower six bits of the subcommutator ID, a value from 0--63, serves as an index to one of the 64 science subcommutator words in the E format (0=$E_{101}$, 63=$E_{232}$.) The actual values of the subcommutator words are stored in the next two 6-bit word positions.
\paragraph{Bits 109--114: Engineering subcommutator} The value of the C-format word indexed by the subcommutator ID. For an explanation of the C format, see the next section.
\paragraph{Bits 115--120: Science subcommutator} The value of the E-format word indexed by the subcommutator ID. For an explanation of the E format, see the next section.

\subsubsection{Data Formats}

The Pioneer 10/11 spacecraft utilized several different formats for telemetry:
\begin{table}[h]
\caption{Science format A bit assignments.\label{tb:formatA}}
\vskip 6pt
\centering
\begin{tabular}{|cl|}
\hline\hline
Bits&Description\\\hline
25--36&GSFC/CRT\\
37--39&UCSD/TRD\\
40--51&GSFC/CRT\\
52--80&UC/CPI\\
81--96&UI/GTT\\
\hline
121--132&GSFC/CRT\\
133--162&JPL/HVM\\
163--171&USC/UV\\
172--192&ARC/PA\\
\hline
\end{tabular}
\end{table}

\paragraph{Science format A} This format was designed for use during the interplanetary cruise. Of each 192-bit telemetry frame, 144 bits were available for use by scientific instruments. The format was shared by seven scientific instruments (Table~\ref{tb:formatA}).

\begin{table}[h]
\caption{Science format B bit assignments.\label{tb:formatB}}
\vskip 6pt
\centering
\begin{tabular}{|cl|}
\hline\hline
Bits&Description\\\hline
25--39&UCSD/TRD\\
40--51&GSFC/CRT\\
52--84&UC/CPI\\
85--96&UI/GTT\\
\hline
121--132&UCSD/TRD\\
133--162&JPL/HVM\\
163--171&USC/UV\\
172--192&ARC/PA\\
\hline
\end{tabular}
\end{table}

\paragraph{Science format B} This format was designed for use during Jupiter encounter. The same seven science instruments as in Format A shared the available 144 data bits in each telemetry frame (Table~\ref{tb:formatB}).

\paragraph{Format C} Formats C1 through C4 contain engineering telemetry words. These telemetry words were six-bit values, transmitted one at a time in a subcommutator field in formats A and B. It was also possible to command the spacecraft to transmit engineering telemetry in accelerated mode; either one of the 4 C formats could be selected, or the spacecraft could be instructed to cycle through the four formats. Note that, as C-format telemetry frames also contained identification and subcommutator information in bits 1--24 and 97--120, C-format words $C_{x01}$--$C_{x04}$ and $C_{x17}$--$C_{x20}$ could never appear in an accelerated frame. For a detailed list of C format telemetry words, see table~\ref{tb:formatC}.

\begin{small}
\begin{longtable}{|p{0.05\linewidth}|c|p{0.7\linewidth}|c|}
\caption{\normalsize Engineering telemetry words (Format C).\label{tb:formatC}}\\
\hline\hline
Word&Bit&Definition&Type\\\hline
\endfirsthead
\caption[]{(continued)}\\\hline
Word&Bit&Definition&Type\\\hline
\endhead
\hline
\endfoot
\hline\hline
\endlastfoot
$C_\mathrm{101}$&&A/D calibration value (low) 0.168~V&A\\
$C_\mathrm{102}$&&A/D calibration value (mid) 1.512~V&A\\
$C_\mathrm{103}$&&A/D calibration value (high) 2.424~V&A\\
$C_\mathrm{104}$&&Extended subcommutator ID.&B\\
$C_\mathrm{105}$&&RTG 2 current (0--11~A)&A\\
$C_\mathrm{106}$&&Battery voltage (0--15~V)&A\\
$C_\mathrm{107}$&&DC bus voltage (26-30~V)&A\\
$C_\mathrm{108}$&1&JPL/HVM power&B\\
&2&ARC/PA power&\\
&3&UC/CPI power&\\
&4&UI/GTT power&\\
&5&GSFC/CRT power&\\
&6&&\\
$C_\mathrm{109}$&&Battery charge current (0--0.3~A)&A\\
$C_\mathrm{110}$&&RTG 1 voltage (0--6~V)&A\\
$C_\mathrm{111}$&&Receiver A AGC CONSCAN $-$4 to +4~dB&A\\
$C_\mathrm{112}$&&Roll attitude timer MSB&B\\
$C_\mathrm{113}$&&RTG 4 voltage (0--6~V)&A\\
$C_\mathrm{114}$&&RTG 3 current (0--11~A)&A\\
$C_\mathrm{115}$&&Battery temperature (--20 to 120$^\circ$)F&A\\
$C_\mathrm{116}$&&Roll attitude timer LSB&B\\
$C_\mathrm{117}$&&TRF +5~V output CDU Bus A (0--6~V)&A\\
$C_\mathrm{118}$&&TRF +5~V output CDU Bus B (0--6~V)&A\\
$C_\mathrm{119}$&&DC bus voltage (0--30~V)&A\\
$C_\mathrm{120}$&1&JPL/HVM bool released (0=released)&B\\
&2&RTG 1/2 deployed (0=deployed)&\\
&3&RTG 3/4 deployed (0=deployed)&\\
&4&Separation status (0=separated)&\\
&5&Decoder A (1=addressed)&\\
&6&Decoder B (1=addressed)&\\
$C_\mathrm{121}$&&Receiver B AGC CONSCAN $-$4 to +4~dB&A\\
$C_\mathrm{122}$&&Shunt bus current (0--3~A)&A\\
$C_\mathrm{123}$&&RTG 4 current (0--11~A)&A\\
$C_\mathrm{124}$&1&UCSD/TRD power&B\\
&2&USC/UV power&\\
&3&UA/IPP power&\\
&4&CIT/IR power&\\
&5&GE/AMD power&\\
&6&LaRC/MD power&\\
$C_\mathrm{125}$&&RTG 2 voltage (0--6~V)&A\\
$C_\mathrm{126}$&&Battery discharge current (0--10~A)&A\\
$C_\mathrm{127}$&&RTG 1 current (0--11~A)&A\\
$C_\mathrm{128}$&1&Battery charge (0=auto, 1=float)&B\\
&2&Battery discharge (0=enabled)&\\
&5&Ordnance relay status prime (1=armed)&\\
&6&Ordnance relay status redundant&\\
$C_\mathrm{129}$&&DC bus current (0--6~A)&A\\
$C_\mathrm{130}$&&Nitrogen tank temperature ($^\circ$F)&A\\
$C_\mathrm{131}$&&RTG 3 voltage (0--6~V)&A\\
$C_\mathrm{132}$&1&Coded/uncoded data (1=coded)&B\\
&2&RTG 1/2 ordnance status (0=safe)&\\
&3&RTG 3/4 ordnance status (0=safe)&\\
&4&RTG 1/2 redundant ordnance status&\\
&5&RTG 3/4 redundant ordnance status&\\
&6&CDU sequencer status (1=enabled)&\\
$C_\mathrm{201}$&&RTG 1 fin root temperature (160--360$^\circ$F)&A\\
$C_\mathrm{202}$&&RTG 2 fin root temperature (160--360$^\circ$F)&A\\
$C_\mathrm{203}$&&RTG 3 fin root temperature (160--360$^\circ$F)&A\\
$C_\mathrm{204}$&&RTG 4 fin root temperature (160--360$^\circ$F)&A\\
$C_\mathrm{205}$&&TWT A temperature (40--125$^\circ$F)&A\\
$C_\mathrm{206}$&&Driver a temperature (20--110$^\circ$F)&A\\
$C_\mathrm{207}$&&TWT A converter temperature (40--125$^\circ$F)&A\\
$C_\mathrm{208}$&&TWT A cathode current (24--30~mA)&A\\
$C_\mathrm{209}$&&Shunt bus current (0--3~A)&A\\
$C_\mathrm{210}$&&Propellant supply pressure (0--600~PSIA)&A\\
$C_\mathrm{211}$&&TWT A helix current (0--10~mA)&A\\
$C_\mathrm{212}$&&Receiver A loop stress (--100 to 100~kHz)&A\\
$C_\mathrm{213}$&&Receiver B signal strength (--149 to --63~dBm)&A\\
$C_\mathrm{214}$&&TWT B RF output power (26--40.4~dBm)&A\\
$C_\mathrm{215}$&&TWT B cathode current (24--30~mA)&A\\
$C_\mathrm{216}$&&TWT B helix current (0--10~mA)&A\\
$C_\mathrm{217}$&&RTG 4 hot junction temperature (880--1200$^\circ$F)&A\\
$C_\mathrm{218}$&&RTG 3 hot junction temperature (880--1200$^\circ$F)&A\\
$C_\mathrm{219}$&&RTG 2 hot junction temperature (880--1200$^\circ$F)&A\\
$C_\mathrm{220}$&&RTG 1 hot junction temperature (880--1200$^\circ$F)&A\\
$C_\mathrm{221}$&&TWT B converter temperature (40--125$^\circ$F)&A\\
$C_\mathrm{222}$&&Receiver A VCO temperature (20--110$^\circ$F)&A\\
$C_\mathrm{223}$&&Driver B temperature (20--110$^\circ$F)&A\\
$C_\mathrm{224}$&&TWT A reference voltage (0--28~V)&A\\
$C_\mathrm{225}$&&+Y PSA line temperature ($^\circ$F)&A\\
$C_\mathrm{226}$&&--Y PSA line temperature ($^\circ$F)&A\\
$C_\mathrm{227}$&&Receiver B VCO temperature (20--110$^\circ$F)&A\\
$C_\mathrm{228}$&&TWT B temperature (40--125$^\circ$F)&A\\
$C_\mathrm{229}$&&Receiver B loop stress (--100 to 100~kHz)&A\\
$C_\mathrm{230}$&&TWT B reference voltage (0--28~V)&A\\
$C_\mathrm{231}$&&TWT A RF output power (26--40.4~dBm)&A\\
$C_\mathrm{232}$&&Receiver A signal strength (--149 to --63~dBm)&A\\
$C_\mathrm{301}$&&Platform temperature 1 (0--140$^\circ$F)&A\\
$C_\mathrm{302}$&&Platform temperature 2 (0--140$^\circ$F)&A\\
$C_\mathrm{303}$&&SRA temperature (--10 to 95$^\circ$F)&A\\
$C_\mathrm{304}$&&Platform temperature 3 (0--140$^\circ$F)&A\\
$C_\mathrm{305}$&&Stored command time register bits 6--1&D\\
$C_\mathrm{306}$&1&Stored command time register bit 0&D\\
&2--6&Stored command register bits 7--3&\\
$C_\mathrm{307}$&1--3&Stored command register bits 2--0&D\\
&4--6&Stored command identification bits 1--3&\\
$C_\mathrm{308}$&1&Receiver A signal present&B\\
&2&Receiver B signal present&\\
&3&Receiver A oscillator enabled&\\
&4&Receiver B oscillator enabled&\\
&5&Spin thruster B pulse count&\\
&6&Spin thruster A pulse count&\\
$C_\mathrm{309}$&&Velocity thruster cluster 1 temperature (40--200$^\circ$F)&A\\
$C_\mathrm{310}$&&Spin thruster cluster temperature (40--200$^\circ$F)&A\\
$C_\mathrm{311}$&&VPT 1 temperature (400--1800$^\circ$F)&A\\
$C_\mathrm{312}$&&VPT 2 temperature (400--1800$^\circ$F)&A\\
$C_\mathrm{313}$&1&CONSCAN thruster phase output status (0=0$^\circ$, 1=180$^\circ$)&D\\
&2&CONSCAN threshold mode status (0=lo, 1=hi)&\\
&3--6&CONSCAN A $\sin{\theta}$ bits 0--3&\\
$C_\mathrm{314}$&1--3&CONSCAN A $\sin{\theta}$ bits 4--6&D\\
&4&CONSCAN A $\sin{\theta}$ sign bit&\\
&5--6&CONSCAN A $\cos{\theta}$ bits 0--1&\\
$C_\mathrm{315}$&1--5&CONSCAN A $\cos{\theta}$ bits 2--6&D\\
&6&CONSCAN A $\cos{\theta}$ sign bit&\\
$C_\mathrm{316}$&1&CONSCAN power&D\\
&2&CONSCAN threshold (1=above)&\\
&4&Receiver switch status (1=A/B=Hi/Med)&\\
&5&Transmitter switch status (1=A/B=Med/Hi)&\\
&6&Antenna feed switch status (1=offset)&\\
$C_\mathrm{317}$&&SSA temperature (--30 to 194$^\circ$F)&A\\
$C_\mathrm{318}$&&Platform temperature 4 (0--140$^\circ$F)&A\\
$C_\mathrm{319}$&&Platform temperature 5 (0--140$^\circ$F)&A\\
$C_\mathrm{320}$&&Platform temperature 6 (0--140$^\circ$F)&A\\
$C_\mathrm{321}$&&Velocity thruster 2 (1B) pulse count&D\\
$C_\mathrm{322}$&&Velocity thruster 4 (2A) pulse count&D\\
$C_\mathrm{323}$&&&\\
$C_\mathrm{324}$&1&Command memory status (0=processing, 1=standby)&B\\
$C_\mathrm{325}$&&VPT 4 temperature (400--1800$^\circ$F)&A\\
$C_\mathrm{326}$&&Velocity thruster cluster 2 temperature (40--200$^\circ$F)&A\\
$C_\mathrm{327}$&&Propellant supply temperature (40--160$^\circ$)&A\\
$C_\mathrm{328}$&&VPT 3 temperature (400--1800$^\circ$F)&A\\
$C_\mathrm{329}$&&Velocity thruster 1 (1A) pulse count&D\\
$C_\mathrm{330}$&&Velocity thruster 3 (2B) pulse count&D\\
$C_\mathrm{331}$&&&\\
$C_\mathrm{332}$&1&Sequencer power&B\\
&2&Overload protection (1=on)&\\
&3&Receiver reverse inhibit&\\
&4&Command processor memory select (1=A)&\\
&5&Command memory DTU identification (1=DTU)&\\
&6&CDU +5~V bus status A/B (1=Bus A)&\\
$C_\mathrm{401}$&&&\\
$C_\mathrm{402}$&&&\\
$C_\mathrm{403}$&1&Precession pair (1=VPT 1\&4)&D\\
&2--4&Pulse length bits 1--3&\\
&5&$\Delta v$ pair (1=VPT 1\&3)&\\
&6&Spin control direction (0=up)&\\
$C_\mathrm{404}$&1--5&Star time gate bits 0--4&D\\
&6&$\Delta v$/SCT mode enabled&\\
$C_\mathrm{405}$&&Spin period MSB&D\\
$C_\mathrm{406}$&&Spin period&D\\
$C_\mathrm{407}$&&Spin period LSB&D\\
$C_\mathrm{408}$&1--5&Roll pulse/index pulse phase error bits 4--0&D\\
&6&Phase error sign (1=roll pulse before index pulse)&\\
$C_\mathrm{409}$&1&VPT 1 firing status&B\\
&2&VPT 2 firing status&\\
&3&VPT 4 firing status&\\
&4&VPT 3 firing status&\\
&5&SCT 1 firing status&\\
&6&SCT 2 firing status&\\
$C_\mathrm{410}$&1&Despin on/off&D\\
&2&CONSCAN enabled&\\
&3&Clock select A&\\
&4&Clock select B&\\
&5&Star angle gate 45$^\circ$/360$^\circ$ (1=45$^\circ$)&\\
&6&Star level $>$180\% Canopus&\\
$C_\mathrm{411}$&&No. 1 precession magnitude bits 0--5&D\\
$C_\mathrm{412}$&1--5&No. 1 precession magnitude bits 6--10&D\\
&6&$\Delta v$ magnitude bit 0&\\
$C_\mathrm{413}$&&$\Delta v$ magnitude bits 1--6&D\\
$C_\mathrm{414}$&&$\Delta v$ magnitude bits 7--12&D\\
$C_\mathrm{415}$&&No. 2 precession magnitude bits 0--5&D\\
$C_\mathrm{416}$&1--5&No. 2 precession magnitude bits 6--10&D\\
&6&Star coincidence&\\
$C_\mathrm{417}$&1&SPSG roll reference (0=0$^\circ$, 1=180$^\circ$)&B\\
&2&SPSG mode bit 1&\\
&3&SPSG mode bit 2&\\
$C_\mathrm{418}$&&&\\
$C_\mathrm{419}$&&Star delay bits 0--5&D\\
$C_\mathrm{420}$&&Star delay bits 6--11&D\\
$C_\mathrm{421}$&1--2&Star count bits 0--2&D\\
&4&CEA power status DSL A&\\
&5&CEA power status DSL B&\\
&6&CEA power status PSE&\\
$C_\mathrm{422}$&6&No. 2 precession redundant magnitude bit 0&D\\
$C_\mathrm{423}$&&No. 2 precession redundant magnitude bits 1--6&D\\
$C_\mathrm{424}$&1&No. 2 precession redundant magnitude bit 7&D\\
&2--6&No. 2 precession angle bits 0--4&\\
$C_\mathrm{425}$&1--4&No. 2 precession angle bits 5--8&D\\
&5--6&$\Delta v$ redundant magnitude bits 0--1&\\
$C_\mathrm{426}$&&$\Delta v$ redundant magnitude bits 2--7&D\\
$C_\mathrm{427}$&&Time delay bits 0--5&D\\
$C_\mathrm{428}$&1&Time delay bit 6&D\\
&2--6&No. 1 precession redundant magnitude bits 0--4&\\
$C_\mathrm{429}$&1--3&No. 1 precession redundant magnitude bits 5--7&D\\
&4--6&No. 1 precession angle bits 0--2&\\
$C_\mathrm{430}$&&No. 1 precession angle bits 3--8&D\\
$C_\mathrm{431}$&1--3&ACS sequence status bits 1--3&D\\
&4&Star time gate enabled&\\
&5--6&Reference select bits 1--2&\\
$C_\mathrm{432}$&1&Star location octant 1 present&\\
&2&Star location octant 8 present&\\
&3&ACS registers inhibit&\\
&4&Precession register 1 arm&\\
&5&$\Delta v$ register 1 arm&\\
&6&Precession register 2 arm&\\
\end{longtable}
\end{small}

\paragraph{Format D} Formats D1 through D3 were special data formats in which all 192 bits in a telemetry frame were assigned to science instruments (i.e., identification, sync, and subcommutator bits were suppressed). The D format frames were telemetered alternately with either an A format or a B format frame, as commanded from the ground. Format D1 was used by the UA/IPP instrument. Format D2 contained data from both the CIT/IR and UA/IPP instruments. Format D3 was reserved for science data from the ARC/PA instrument.

\paragraph{Format E} Format E was reserved for science instrument housekeeping telemetry words. E format words appeared only in the subcommutator of A format or B format frames. For a detailed listing of E-format telemetry words, see table~\ref{tb:formatE}.

\begin{small}
\begin{longtable}{|p{0.05\linewidth}|c|p{0.7\linewidth}|c|}
\caption{\normalsize Science instrument telemetry words (Format E).\label{tb:formatE}}\\
\hline\hline
Word&Bit&Definition&Type\\\hline
\endfirsthead
\caption[]{(continued)}\\\hline
Word&Bit&Definition&Type\\\hline
\endhead
\hline
\endfoot
\hline\hline
\endlastfoot
$E_\mathrm{101}$&&ARC/PA detectors temperature&A\\
$E_\mathrm{102}$&&ARC/PA electronics temperature&A\\
$E_\mathrm{103}$&&JPL/HVM spectrum analyzer X-axis output&A\\
$E_\mathrm{104}$&&JPL/HVM spectrum analyzer Y-axis output&A\\
$E_\mathrm{105}$&&JPL/HVM spectrum analyzer Z-axis output&A\\
$E_\mathrm{106}$&&JPL/HVM status&D\\
$E_\mathrm{107}$&&LaRC/MD event count&\\
$E_\mathrm{108}$&1&UC/CPI detector 1 status&B\\
&2&UC/CPI detector 2 status&\\
&3&UC/CPI detector 7 status&\\
&4&UC/CPI priority mode status&\\
&5&UC/CPI calibrate status&\\
&6&UC/CPI calibrate status&\\
$E_\mathrm{109}$&&USC/UV electronics temperature&A\\
$E_\mathrm{110}$&&UC/CPI electronics temperature&A\\
$E_\mathrm{111}$&&UC/CPI egg current range 1&A\\
$E_\mathrm{112}$&&UC/CPI egg current range 2&A\\
$E_\mathrm{113}$&&UC/CPI egg current range 3&A\\
$E_\mathrm{114}$&&UC/CPI fission detector&D\\
$E_\mathrm{115}$&&UC/CPI fission detector&D\\
$E_\mathrm{116}$&&UC/CPI fission detector&D\\
$E_\mathrm{117}$&&CIT/IR low range temperature&A\\
$E_\mathrm{118}$&&GE/AMD preamp temperature&A\\
$E_\mathrm{119}$&&GE/AMD secondary voltage&A\\
$E_\mathrm{120}$&&&\\
$E_\mathrm{121}$&&&\\
$E_\mathrm{122}$&&GE/AMD event data&D\\
$E_\mathrm{123}$&&GE/AMD event data&D\\
$E_\mathrm{124}$&1&UI/GTT logic status&B\\
&2&GSFC/CRT status&\\
&3&GE/AMD star exclusion status&\\
&4&GE/AMD data readout status&\\
&5&USC/UV channel status&\\
&6&USC/UV roll status&\\
$E_\mathrm{125}$&&GSFC/CRT electronics temperature&A\\
$E_\mathrm{126}$&&GSFC/CRT analog data 1&A\\
$E_\mathrm{127}$&&GSFC/CRT analog data 2&A\\
$E_\mathrm{128}$&&GSFC/CRT detector temperature&A\\
$E_\mathrm{129}$&&GSFC/CRT secondary voltage&A\\
$E_\mathrm{130}$&&GSFC/CRT identification data&D\\
$E_\mathrm{131}$&&CIT/IR command register part 1&D\\
$E_\mathrm{132}$&&CIT/IR command register part 2&D\\
$E_\mathrm{201}$&&CIT/IR high range temperature&A\\
$E_\mathrm{202}$&&JPL/HVM commutated housekeeping data&A\\
$E_\mathrm{203}$&&JPL/HVM spectrum analyzer X-axis output&A\\
$E_\mathrm{204}$&&JPL/HVM spectrum analyzer Y-axis output&A\\
$E_\mathrm{205}$&&JPL/HVM spectrum analyzer Z-axis output&A\\
$E_\mathrm{206}$&&UC/CPI L1 PHA&D\\
$E_\mathrm{207}$&&LaRC MD event count&D\\
$E_\mathrm{208}$&1&GE/AMD threshold level status&B\\
&2&GE/AMD bandwidth status&\\
$E_\mathrm{209}$&&UCSD/TRD electronics temperature&A\\
$E_\mathrm{210}$&&UCSD/TRD high voltage monitor&A\\
$E_\mathrm{211}$&&UCSD/TRD PCU monitor&A\\
$E_\mathrm{212}$&&UC/CPI D7 count rate&A\\
$E_\mathrm{213}$&&UC/CPI egg temperature&A\\
$E_\mathrm{214}$&&UC/CPI L1 L2 coincidence count rate&D\\
$E_\mathrm{215}$&&UC/CPI D1 $\overline{\mathrm{S}}$ D3 $\overline{\mathrm{D}}$7 count rate&D\\
$E_\mathrm{216}$&&UC/CPI D2 D4 D5 D6 $\overline{\mathrm{D}}$7 count rate&\\
$E_\mathrm{217}$&&UA/IPP high voltage monitor&\\
$E_\mathrm{218}$&&&\\
$E_\mathrm{219}$&&&\\
$E_\mathrm{220}$&&UI/GTT 7.75~V monitor&A\\
$E_\mathrm{221}$&&UI/GTT electronics temperature&A\\
$E_\mathrm{222}$&&GE/AMD event data&D\\
$E_\mathrm{223}$&&GE/AMD event data&D\\
$E_\mathrm{224}$&2&UCTD/TRD timing status&B\\
&3&UCTD/TRD high voltage status&\\
&4&UC/CPI GSE stimulus status&\\
&5&USC/UV channel status&\\
&6&USC/UV roll status&\\
$E_\mathrm{225}$&&&\\
$E_\mathrm{226}$&&&\\
$E_\mathrm{227}$&&&\\
$E_\mathrm{228}$&&&\\
$E_\mathrm{229}$&&&\\
$E_\mathrm{230}$&&&\\
$E_\mathrm{231}$&&&\\
$E_\mathrm{232}$&&&\\
\end{longtable}
\end{small}

\subsubsection{Calibration coefficients}

Telemetry data words in telemetry frames contained analog, digital, or binary information. Analog information included sensor readings, such as voltages, currents, or temperatures. Digital telemetry values represented counters and timers. Binary telemetry readings included switch, sensor, and logic states.

Decoding digital and binary values is relatively trivial, once telemetry word assignments are known. Decoding analog values is trickier. Analog readings were stored in the form of 6-bit words, representing 64 analog levels.

The analog/digital converter measured voltages in 48~mV increments. Level 0 was reserved for voltages below 0~V (down to a minimum of --10~V); level 63 represented voltages above 2.976~V (up to a maximum of 5~V). The response of the A/D converter was strictly linear.

Each analog sensor on board each spacecraft was individually calibrated before launch and precise values of the analog-to-digital conversion were determined and fitted to a fifth-order polynomial:
\begin{equation}
x_\mathrm{analog}=A_5x_\mathrm{digital}^5+A_4x_\mathrm{digital}^4+A_3x_\mathrm{digital}^3+A_2x_\mathrm{digital}^2+A_1x_\mathrm{digital}+A_0.
\end{equation}

The calibration procedure also determined calibration limits for each sensor.

Calibration information for both Pioneer 10 and 11 survived in the form of LabView program code written by Larry Kellogg \cite{LK031130}. Documentation also exists containing calibration information, limits, and charts for Pioneer 10 \cite{ARC037}. The data found in these two sources are consistent.

\subsection{Media and Data Recovery}

Originally, Pioneer 10/11 telemetry was stored on magnetic tape. To our great fortune, Pioneer telemetry tapes from the 1970s and the 1980s were copied to magneto-optical (``floptical'') data cartridges, during a project that was initiated in the early 1990s. This not only saved the data from eventual planned destruction, but also kept the data readable; it is highly doubtful that any surviving 30-year old Pioneer magnetic tapes are readable today, even if some could still be located.

The ``floptical'' cartridges had a capacity of 128~MB each. One or more ``floptical'' cartridges were used to store each year's worth of data for each spacecraft. Early in the missions, when DSN coverage of the spacecraft was near continuous and a high data rate was used for the transmission, a the annual data volume far exceeded the capacity of a single cartridge; for instance, data for Pioneer 10 for 1972 was stored on 36 cartridges. Later, much fewer cartridges were required; in the late 1980s, early 1990s, when the spacecraft were already in the outer solar system, only 2-3 cartridges a year could store all the data that was received.

\begin{table}
\caption{Pioneer 10/11 missing MDRs (periods of missing data shorter than 1-2 days not shown.)\label{tb:missing_mdr}}
\vskip 6pt
\centering
\begin{tabular}{|l|l|l|}\hline
Spacecraft&Year&DOYs\\\hline\hline
Pioneer-10&1972&133-149\\
&1973&004-008, 060-067, 332-341\\
&1974&034-054\\
&1979&025-032, 125-128, 137-157, 171-200\\
&1980&173-182, 187-199, 248-257\\
&1983&329-348\\
&1984&346-359\\\hline
Pioneer-11&1973&056-064, 067-080, 082-086, 088-094\\
&1980&309-330, 337-365\\
&1982&318-365\\
&1983&001-050\\
&1984&343-357\\
&1990&081-096\\\hline\hline
\end{tabular}
\end{table}

The transcription from magnetic tape to ``floptical'' cartridges was documented. We have hand-written log sheets containing information about each tape that was transcribed. From these logs, we can determine what data files were missing and for what reason: in some cases, both the primary and backup tapes were unreadable, while in other cases, the tapes were missing (see Table~\ref{tb:missing_mdr}).

The total number of MDR data files is 8,976 (in 160 directories) for Pioneer 10, and 7,540 (in 218 directories) for Pioneer 11. They contain 16.33~GB and 23.01~GB, respectively\footnote{It may appear odd at first that there is more data for Pioneer 11 even though its mission duration was shorter. The explanation lies in the fact that Pioneer 11 had an encounter with Saturn, and in the period before, during, and after Saturn encounter, much data was telemetered at high speed.}.

Curiously, the Pioneer 11 data set contains records that predate the launch of the spacecraft; the data contained in these records suggests that these MDRs may have been collected during pre-launch tests.

It is possible to compress all telemetry for one spacecraft using standard compression tools ({\tt gzip}) so that it fits on one standard DVD; alternatively, data for both spacecraft can be made to fit on a single double-layer DVD.

\subsection{Data reliability}

Thousands of files containing tens of gigabytes of data, stored on magnetic tape for years, copied over to other obsolete media formats, and then once again transmitted through network interfaces and copied to a personal computer... just how reliable is this data set? Can it at all be trusted?

And perhaps equally importantly, was the data reliable in the first place? Even before it was converted to MDR format, the telemetry frames underwent many steps of processing as they were received by the DSN, packaged, and re-packaged.

An examination of the data set provides answers to these questions.

First, it appears unlikely that there were any significant errors during transcription between various media formats. The data does not appear corrupt: it is structurally intact, with MDR and telemetry frames clearly identifiable and containing expected values in fixed fields.

Second, slowly changing values (e.g., temperature, spin rate) follow expected trends and can be compared against other project documentation. This reassures us that the values extracted from the files are valid.

Having said that, the files contain a significant number of ``bad'' records. These records have valid MDR header and footer fields, but bogus information in place of the telemetry frame. The number of bad records increases towards the end of the mission. This clearly suggests that these records were a result of bad reception by the DSN of an extremely weak spacecraft signal, and not transcription errors as the received data was moved between different media types.

We find that despite these errors, it is possible to extract highly reliable readings for most telemetry values. For instance, for slowly changing values we often employed an algorithm that retrieved all readings for that value within a given time period, and used a ``majority rule'' algorithm to decide which value to use. We found that such an extraction resulted in very ``clean'' data, with data sets of tens of thousands of readings containing only a handful of bad records. Moreover, the bad records were clearly identifiable ``by hand'', as they contained completely bogus data (e.g., a single reading showing thrice the normal spin rate.)

Moreover, when there was a possibility to independently verify a reading (e.g., by comparing it with values computed from first principles, or comparing redundant readings against each other) the results were always as expected. This not only suggests that the data is good, it also confirms that, for analog readings, the calibrated digital/analog conversion factors were valid.

\section{Software}
\label{sec:tools}

In order to be able to produce the data products described in the previous section, and also to aid our investigation of the Pioneer anomaly by providing a means for ad-hoc data retrieval, we developed several software tools that we used to access and process Pioneer 10/11 MDRs.

These tools can be categorized as follows:
\begin{itemize}
\item A software library to be used by other programs for accessing MDRs;
\item A command-line tool for expert MDR processing;
\item Interactive application for viewing MDRs;
\item Interactive interfaces for convenient ad-hoc data retrieval;
\item Command-line scripts for generating data product files.
\end{itemize}

\subsection{Low-level MDR data extraction library}

The low-level extraction library is a C-language library providing a small set of functions, callable from programs written in C or other languages that support the concept of callback functions, for accessing MDRs.

The software library implements MDR retrieval in a loop. The loop is internal to the library code, but it calls a user-supplied callback function whenever the desired parameter is located. The library iterates through all MDR content until either it is signaled to stop or the MDR set is exhausted.

The main MDR retrieval function is declared as

\vskip 12pt
\begin{tabular}{|p{0.9\linewidth}|}\hline
{\tt bool GetMDR(int nSCID, time\_t tStamp, const char *pszField,}\\
{\tt ~~~~~~~~~~~~int nSkip, bool bFilter, GETMDRCALLBACKFN pfnGetMDRCallBack);}\\
\hline
\end{tabular}
\vskip 12pt

where the function parameters are defined as follows:

\begin{description}
\item {\tt nSCID} is the spacecraft identifier (23 for Pioneer 10, 24 for Pioneer 11);
\item {\tt tStamp} is the start time stamp (in UNIX {\tt time\_t} format) of the first data point to be retrieved;
\item {\tt pszField} is the identifier of the field to be retrieved. It is a character string in the form of {\tt "C-$nnn$} or {\tt "E-$nnn$}, where $nnn$ is three decimal digits from 101 through 432 (for C words) or 101 through 232 (for E words);
\item {\tt nSkip} is the number of MDRs to skip between each read. The library normally reads each MDR looking for the desired parameter (i.e., {\tt nSkip = 1}.) When {\tt nSkip} is set to $n$ ($n>1$), only every $n^\mathrm{th}$ record is accessed in the binary MDR files;
\item {\tt bFilter} determines whether or not the software is to attempt to filter ``bad'' records. In the present implementation, when {\tt bFilter} is set to {\tt true}, MDRs in which the record type in the MDR header and in the telemetry frame do not agree are excluded.
\item {\tt pfnGetMDRCallBack} is a user-defined callback function that is called once each time the specified parameter is found.
\end{description}

The return value of {\tt GetMDR} is {\tt true} if the retrieval was successful, {\tt false} if no data was retrieved.

The {\tt pfnGetMDRCallBack} is declared as follows:

\vskip 12pt
\begin{tabular}{|p{0.9\linewidth}|}\hline
{\tt typedef}\\
{\tt ~~~~bool (*GETMDRCALLBACKFN)(const char *pszName, int nPos,}\\
{\tt ~~~~~~~~~~~~~~~~~~~~~~~~~~~~~time\_t tStamp, int nmSec, int nDQI, int nBPS,}\\
{\tt ~~~~~~~~~~~~~~~~~~~~~~~~~~~~~const char *pszField, const char *pszUnit,}\\
{\tt ~~~~~~~~~~~~~~~~~~~~~~~~~~~~~const char *pszDesc, bool bSubComm,}\\
{\tt ~~~~~~~~~~~~~~~~~~~~~~~~~~~~~int nDValue, const char *pszAValue);}\\
\hline
\end{tabular}
\vskip 12pt

where the function parameters are defined as follows:

\begin{description}
\item {\tt pszName} is the name of the currently processed MDR file;
\item {\tt nPos} is the byte position of the current record in the current MDR file;
\item {\tt tStamp} is the time stamp (in UNIX {\tt time\_t} format) of the current record;
\item {\tt nmSec} is the millisecond part of the timestamp of the current record;
\item {\tt nDQI} is the data quality indicator of the current record, extracted from the MDR frame;
\item {\tt nBPS} is the telemetry data rate, extracted from the MDR frame;
\item {\tt pszField} is the name of the field that is being retrieved (not presently used, set to {\tt NULL};
\item {\tt pszUnit} is the physical unit of measurement (if any) associated with the field being retrieved;
\item {\tt pszDesc} is the verbatim description of the field being retrieved;
\item {\tt bSubComm} is set to {\tt true} when the field was found in the subcommutator, {\tt false} if it was located in an accelerated engineering frame;
\item {\tt nDValue} is the ``digital'' telemetry value;
\item {\tt pszAValue} is the decoded telemetry value; for analog telemetry fields, this parameter points to a string containing the converted (floating point) value. For values representing bit fields, this parameter points to a string containing a 6-digit binary number. In both cases the pointer points to a buffer allocated internally by the library; applications should not manipulate this buffer, only read the data provided through it in the form of a null-terminated string.
\end{description}

The return value of this user-defined callback function should be {\tt true} if more data is desired, {\tt false} otherwise.

The low-level library provides a set of additional helper functions for application programs:

\vskip 12pt
\begin{tabular}{|p{0.9\linewidth}|}\hline
{\tt bool GetInfo(int nSCID, const char *pszField, char **ppszDesc,}\\
{\tt ~~~~~~~~~~~~~char **ppszUnit);}\\
\hline
\end{tabular}
\vskip 12pt
retrieves the description and physical unit name for a given telemetry value. The function is called with the following parameters:
\begin{description}
\item {\tt nSCID} is the spacecraft identifier (23 for Pioneer 10, 24 for Pioneer 11);
\item {\tt pszField} is a character string containing the field name (e.g., {\tt "C-223"});
\item {\tt ppszDesc} is a pointer to a character pointer that will receive a pointer to the field description;
\item {\tt ppszUnit} is a pointer to a character pointer that will receive a pointer to the physical unit associated with the field.
\end{description}
NB: The character buffers pointed to by {\tt *ppszDesc} and {\tt *ppszUnit} are statically allocated and must not be manipulated or overwritten by application programs.

The return value of {\tt GetInfo} is {\tt true} in case of success, {\tt false} if an error occurred.

\vskip 12pt
\begin{tabular}{|p{0.9\linewidth}|}\hline
{\tt const char *GetCFMT(int nValue, int n, int i, bool bP11,}\\
{\tt ~~~~~~~~~~~~~~~~~~~~bool bSubFrame = false);}\\
{\tt const char *GetEFMT(int nValue, int n, int i, bool bP11,}\\
{\tt ~~~~~~~~~~~~~~~~~~~~bool bSubFrame = false);}\\
\hline
\end{tabular}
\vskip 12pt
translate raw (binary) telemetry values into decoded and formatted character strings. For C words, use {\tt GetCFMT}; for E words, use {\tt GetEFMT}. The synopsis for the two functions is identical. Function parameters are as follows:
\begin{description}
\item {\tt nValue} is the digital (undecoded) parameter value;
\item {\tt n} is the number of the telemetry page (i.e., the first digit of the telemetry word index; 1 through 4 for C words, 1 through 2 for E words);
\item {\tt i} is the telemetry word index (1--32);
\item {\tt bP11} is set to {\tt false} for Pioneer 10, {\tt true} for Pioneer 11, and determines which set of calibration tables to use for analog value decoding;
\item {\tt bSubFrame} is to be set to true if the telemetry value was obtained from a subcommutator; if it is a ``main frame only'' telemetry value, the function will decode it as {\tt "----"} indicating the absence of a valid reading.
\end{description}
The return value of these functions is a pointer to an internally allocated character buffer containing the translated value. The buffer should not be overwritten or otherwise manipulated by application programs.

These low-level library functions have been implemented for both 32-bit UNIX and Windows target platforms.

\subsection{Command-line MDR processing}

The library subroutines described in the previous section can be put to immediate use through a command line tool, {\tt mdrread}. This program provides quick and convenient access to MDR data, and it is especially suitable for use in conjunction with UNIX command-line based text processing tools.

The synopsis of {\tt mdrread} is as follows:

\vskip 12pt
\begin{tabular}{|p{0.9\linewidth}|}\hline
{\tt mdrread [-c] [-f] [-l] [-r] [-d datadir] SCID parameter}\\
{\tt ~~~~~~~~timestamp1 timestamp2 [skip]}\\
\hline
\end{tabular}
\vskip 12pt
where
\begin{description}
\item {\tt -c} instructs the program to print an output line only for changed values;
\item {\tt -f} instructs the program to filter bad records from its output;
\item {\tt -l} prevents the program from stopping when too many records ($>10000$) or retrieved or the program has run in excess of 3 minutes;
\item {\tt -r} instructs the program to filter values that fall outside calibrated ranges;
\item {\tt -d datadir} specifies the location of the MDR data directory (default is the current directory; the data directory is supposed to contain subdirectories with names such as {\tt 23P7201}, in which are files named like {\tt m2372063.mdr});
\item {\tt SCID} is the spacecraft identifier (23 for Pioneer 10, 24 for Pioneer 11);
\item {\tt parameter} is the name of the telemetry value to retrieve (e.g., {\tt C-223});
\item {\tt timestamp1} is the UNIX {\tt time\_t} style timestamp of the start date for retrieval;
\item {\tt timestamp2} is the UNIX {\tt time\_t} style timestamp of the end date for retrieval;
\item {\tt skip} is the numeric value for the number of MDR records to skip between retrievals.
\end{description}

By way of example, here is how one would retrieve the values of parameter $C_\mathrm{223}$ for Pioneer 10, for the one minute period between 16:00 and 16:01 on March 3, 1972:

\begin{verbatim}
$ mdrread -d /data/PIONEER/ 23 C-223 `date -u -d "1972-03-03 16:00:00" +%s`\
 `date -u -d "1972-03-03 16:01:00" +%s`
C-223 Driver B Temperature (ºF)
Pathname/Filename       Pos     Timestamp       DQI     SubCom  Binary  Decoded
23P7201/m2372063.mdr    54030   68486406.054    3       Y       15      86.606
23P7201/m2372063.mdr    54062   68486429.830    3       Y       15      86.606
23P7201/m2372063.mdr    54094   68486453.863    3       Y       15      86.606
.end
\end{verbatim}

This example demonstrates how UNIX-style timestamps can be easily generated on a UNIX system using the {\tt date} command. The {\tt date} command can be instructed to turn a user-supplied date ({\tt -d} option) in UTC ({\tt -u} option) into a {\tt time\_t} style timestamp ({\tt +\%s}).

This example also demonstrates how {\tt mdrread} outputs its results. The first line of its output contains the parameter identifier, parameter name, and in parenthesis, the physical unit associated with that parameter. The next line is a header line, followed by data lines; these lines each are divided into seven tab-delimited columns containing the current MDR file name, record position, timestamp (UNIX {\tt time\_t} style value augmented with 3 decimal digits representing milliseconds), the data quality indicator, a yes/no type field identifying whether or not the value was found in a subcommutator, and lastly, the binary and decoded values of the parameter.

UNIX style timestamps can be turned back into human readable days in several ways. If the data is imported into Microsoft Excel, for instance, one can use a formula like {\tt =A1/86400+25569}, which translates a UNIX-style timestamp in cell {\tt A1} into an Excel-style date/time value. If the data is to be further processed by UNIX text processing tools, we provide an ancillary command-line tool, {\tt undate}, for this purpose:

\begin{verbatim}
$ undate 68486406
Fri Mar  3 16:00:06 1972
\end{verbatim}

\subsection{Interactive MDR viewer for Microsoft Windows}

We also developed a more elaborate Windows-based application for viewing MDRs. This application is especially useful when one wishes to browse the contents of an MDR file, for instance, in order to determine the degree to which data might have been corrupted.

\begin{figure}
\centering
\psfig{figure=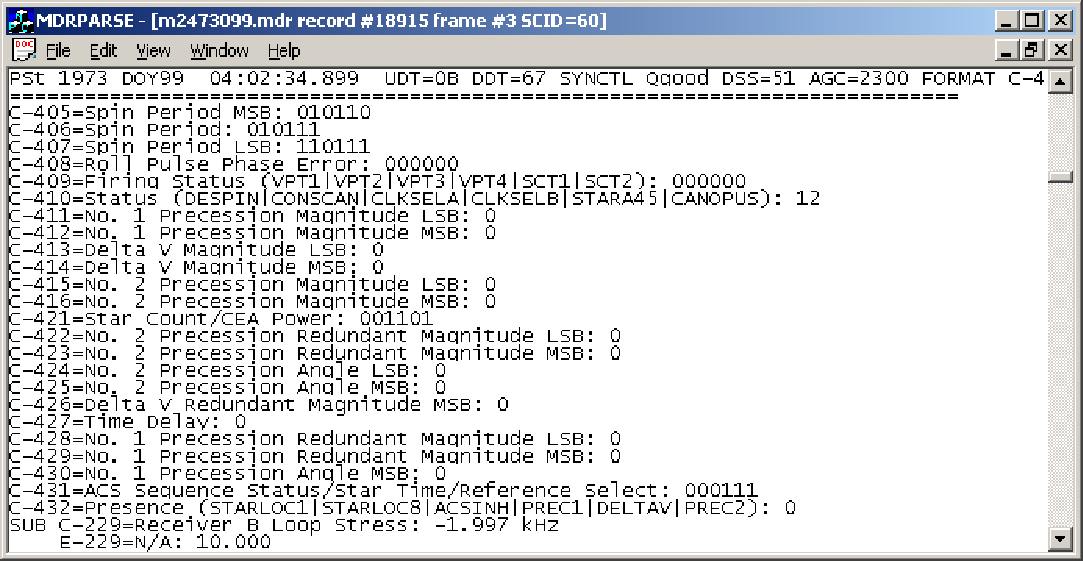,scale=0.5}
\caption{\noindent The interactive Windows application for reading MDRs.\label{fig:mdrparse}}
\end{figure}

The application, depicted in Figure~\ref{fig:mdrparse}, has a very simple user interface. It uses the Windows File Open dialog to let the user select an MDR file, and then it displays the file's decoded contents. The program recognizes all telemetry frame types; however, detailed contents are displayed only for accelerated engineering (C-format) frames.

The application also lets the user copy the visible data area to the Windows clipboard, in the form of ASCII text. Printing of a file, or portions of a file, is supported as well.

\subsection{Ad-hoc data retrieval through an HTML interface}

Though the interactive Windows application was a very useful tool to become familiar with the contents of available MDR files, it is not a very practical tool when one's objective is to extract a parameter for further processing. This is better accomplished using the {\tt mdrread} command-line tool, but that tool may be cumbersome to use. As a third alternative, we developed a set of HTML pages, viewable in any standards-compliant Web browser, that make it easy to access and retrieve telemetry data on an ad-hoc basis.

\begin{figure}
\centering
\psfig{figure=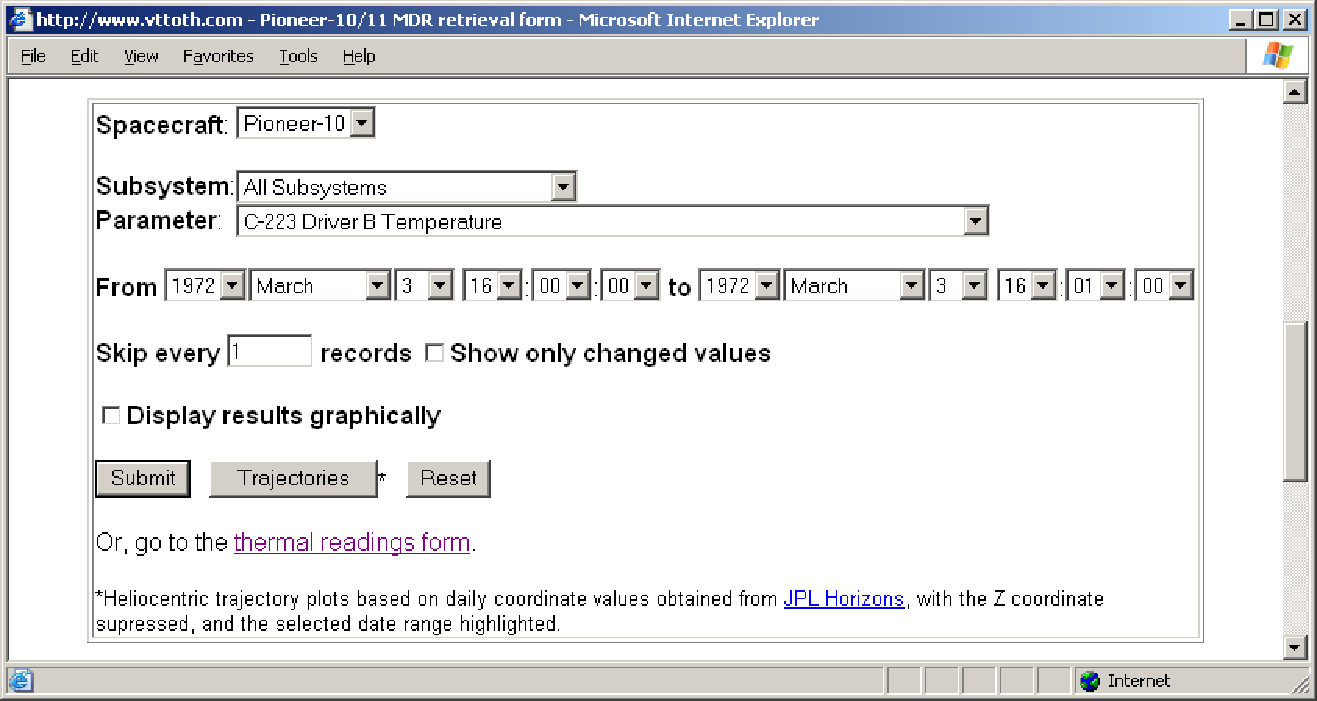,scale=0.5}
\caption{\noindent Interactive Web page for reading MDRs.\label{fig:masterform}}
\end{figure}

The first of these HTML pages, shown in Figure~\ref{fig:masterform}, provides a form-based interface through which any of the C or E-words can be selected for retrieval. The user can specify the spacecraft, the parameter, and a time interval; the result is presented in the form of tab-delimited ASCII text, which most browsers can readily display.

The application engine is written in the form of server-side program code in the PHP and Perl languages. At present, the application back-end is tested to run only using the Apache Web server on a Linux operating system platform.

In addition to ASCII text, the HTML page can also be used to retrieve the data graphically. This option is most useful for analog values, such as thermal or electrical readings.

The HTML interface also presents a third option, the option to retrieve trajectory data, in graphical form, for the Pioneer spacecraft in the solar system. This simple feature uses solar system ephemeris data obtained from JPL's public servers.

\begin{figure}
\centering
\psfig{figure=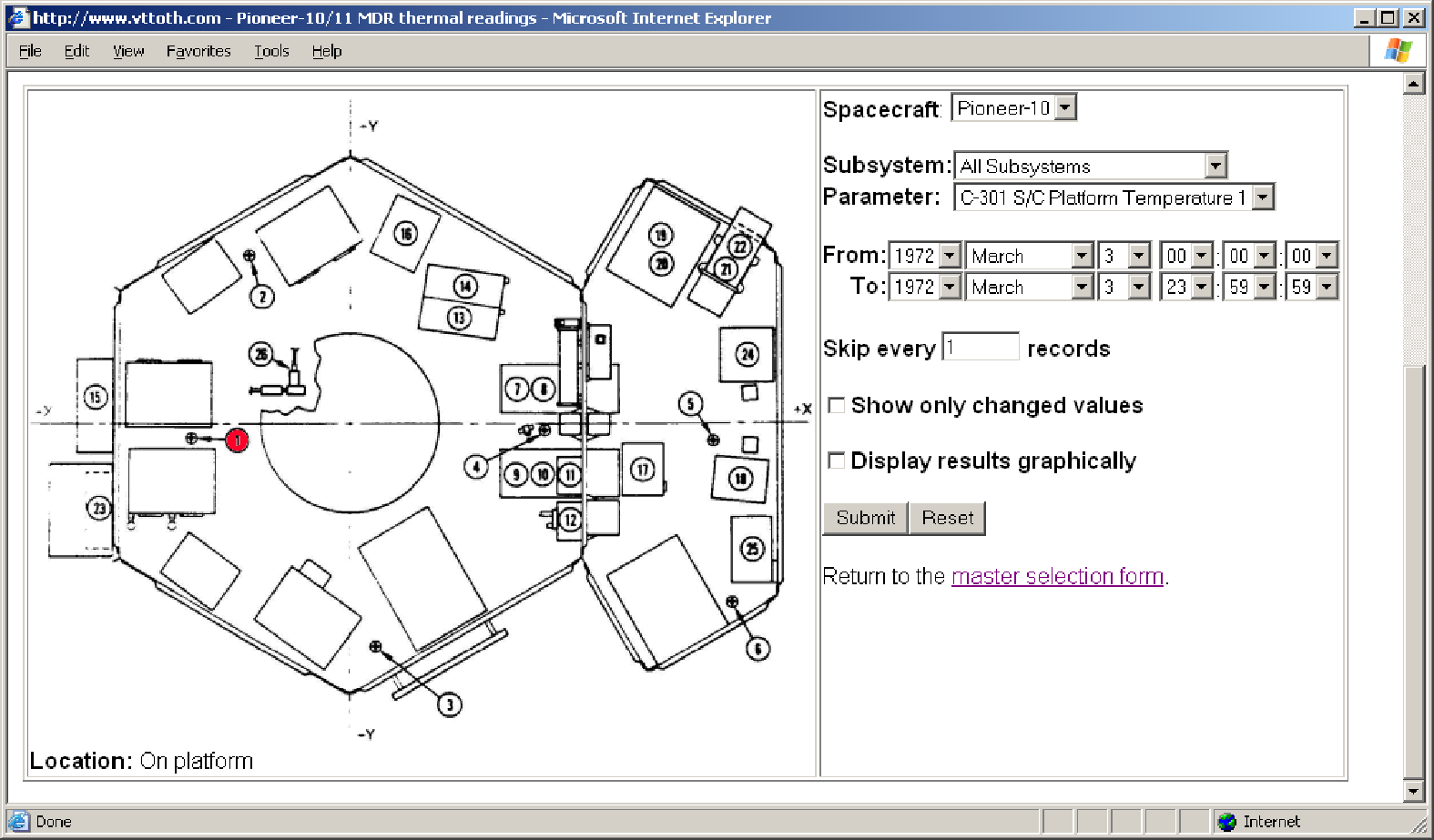,scale=0.5}
\caption{\noindent Interactive Web page for reading MDRs.\label{fig:thermalform}}
\end{figure}

A second Web page, shown in Figure~\ref{fig:thermalform}, provides essentially the same functionality, but restricted for temperature readings only. The interface is enhanced with a diagram of the spacecraft body, with the location of temperature sensors identified. This makes it possible for the user to easily locate a sensor of interest, simply by clicking the image and then selecting the sensor from a dropdown list that contains only the list of sensors at the location where the click took place.

Even with this ad-hoc interface, retrieving parameters can be cumbersome. That is due to the fact that the low-level MDR library is not very efficient; if the user selects a long timespan, it can take an extremely long time (hours or more) for the procedure to finish. In order to avoid such lengthy processing, the retrieval program is configured by default to halt when it retrieved 10,000 records or 3 minutes of processing time; this, however, is far from sufficient to characterize the behavior of a parameter throughout the entire mission duration. The remedy is to use a large ``skip'' parameter to reduce the number of records that the programs need to process; this, however, is also not a perfect solution, as a large ``skip'' value may be appropriate during the initial, data-rich phases of the mission, but not appropriate near the end when data was scarce. In any case, such processing can be very demanding on the server where the data resides.

To remedy this situation, we prepared a large number of plots that are available for viewing offline. Presented in both thumbnail and full-size form, these plots make it very easy for a user to understand the temporal behavior of a parameter, and perhaps zero in on a region of interest. Two HTML pages, containing thumbnails for Pioneer 10 and 11, respectively, are presented for this purpose. For each parameter, the pages contain two links, one to the full-size plot image, one to a tab-delimited ASCII file that contained the data used to generate the image.

\subsection{Data extraction scripts}

Once we were familiar with the data, our objective became clear: we needed to extract data sets containing specific parameter values, characterized for the entire duration of the Pioneer 10/11 missions. Our tools were up to the task, but we still needed a way to accomplish this task with a minimum effort, and in a consistent and reproducible manner.

For this purpose, we developed a series of application scripts, written in UNIX shell ({\tt bash}) language. The scripts had many common characteristics. First, each script established the start date, stop date, and spacecraft identifier in the form of shell variables:

\begin{verbatim}
    STARTDATE=`date -u -d "1972-03-01 00:00:00" +%s`
    ENDDATE=`date -u -d "2002-12-31 23:59:59" +%s`
    CRAFT=23
\end{verbatim}

The scripts used a common subroutine to extract hourly values for any parameter. This subroutine used UNIX command-line text processing tools to process the output of the {\tt mdrread} command:

\begin{verbatim}
    getpar()
    {
      NAME=C$1
      eval ${NAME}=`nice ./mdrread -d /data/PIONEER/ ${CRAFT} C-$1 ${DATE} \
           $((DATE+3599)) | cut -f 7 | uniq -c | sort -n | tail -1 | cut -c 9-`
      if [[ ${!NAME:1:1} > "9" ]] ; then
        eval ${NAME}=""
      fi
    }
\end{verbatim}

with these prerequisites in place, the main loop of the script could begin, performing a function similar to the following example, which would retrieve the fin root temperatures for all four RTGs:

\begin{verbatim}
    for ((DATE=STARTDATE; DATE<ENDDATE; DATE+=3600)) ; do
      getpar 201
      getpar 202
      getpar 203
      getpar 204
      if [ "${C201}${C202}${C203}${C204}" != "" ] ; then
        echo -e "`./undate ${DATE}`\t${C201}\t${C202}\t${C203}\t${C204}
      fi
    done
\end{verbatim}

In some cases, the values were further processed by the script. For the electrical subsystem, for instance, specific mathematical expressions were applied to compute desired electrical values. For the thruster pulse counts, a different logic was used, to ensure that only {\it changes} in the pulse count were recorded, as it is a change in pulse count that signifies a thruster event.

The specific scripts we developed correspond with the data products that we wished to produce:

\vskip 12pt
\begin{tabular}{rl}
{\tt 23comm.sh}:&Communication subsystem parameters for Pioneer 10\\
{\tt 24comm.sh}:&Communication subsystem parameters for Pioneer 11\\
{\tt 23elec.sh}:&Electrical system parameters for Pioneer 10\\
{\tt 24elec.sh}:&Electrical system parameters for Pioneer 11\\
{\tt 23prop.sh}:&Propulsion system parameters for Pioneer 10\\
{\tt 24prop.sh}:&Propulsion system parameters for Pioneer 11\\
{\tt 23pulse.sh}:&VPT pulse counts for Pioneer 10\\
{\tt 24pulse.sh}:&VPT pulse counts for Pioneer 11\\
{\tt 23spin.sh}:&Spin rate measurements for Pioneer 10\\
{\tt 24spin.sh}:&Spin rate measurements for Pioneer 11\\
{\tt 23temp.sh}:&Platform and RTG temperatures for Pioneer 10\\
{\tt 24temp.sh}:&Platform and RTG temperatures for Pioneer 11\\
{\tt 23ttemp.sh}:&Thruster and thruster cluster temperatures for Pioneer 10\\
{\tt 24ttemp.sh}:&Thruster and thruster cluster temperatures for Pioneer 11\\
\end{tabular}

\section{Data products}
\label{sec:data}

A binary data set in a nonstandard format occupying nearly 40~GB of storage space is an unwieldy thing. Although we have developed tools for ad-hoc data retrieval, these tools are not very efficient; it takes many hours, for instance, to retrieve all readings for a given parameter for an entire mission duration. Furthermore, the result of such retrieval still needs to be hand-edited to remove bad data.

For this reason, we prepared several data products that contain data extracted from the telemetry and hand-edited using a consistent editing strategy.

The selection of data sets was guided by our desire to use Pioneer 10/11 in the investigation of the Pioneer anomaly. We concentrated on electrical, thermal, communication, and propulsion system information. We were interested in data that represented overall spacecraft performance, not the performance or internal readings of a specific instrument.

Accordingly, we prepared a set of seven data files for each spacecraft. These data files are either in a tab-delimited ASCII text format, or they are Microsoft Excel format spreadsheets. The spreadsheet format was used for data products that incorporate not only telemetry information but information from other sources as well.

The ASCII format files can also be imported into Microsoft Excel. Text format dates can be converted to Microsoft Excel date/time values using the following Excel formula:
\begin{verbatim}
    =DATEVALUE(CONCATENATE(TRIM(MID(A1,9,2)),"-",MID(A1,5,3),"-",MID(A1,21,4)))+
     TIMEVALUE(MID(A1,12,8))
\end{verbatim}

The ASCII format text files can also be plotted using {\tt gnuplot}. Before plotting, however, it may be necessary to replace consecutive tab characters (indicating a missing value) with two tab characters separated by a bogus value (e.g., the letter X), in order to ensure that {\tt gnuplot} plots the columns correctly, including missing data. Afterwards, a column can be plotted using a {\tt gnuplot} script like the following:
\begin{verbatim}
    set xdata time
    set timefmt '%b %d %H:%M:%S %Y'
    set format x '%Y'
    plot 'file' every ::3 using 2:n
\end{verbatim}
where {\tt n} is the number of the column to be plotted, starting with 6.

\subsection{Common editing strategy}

All data files were produced using an identical editing strategy. First, our command-line scripts were used to produce a ``raw'' version of the data using one-hour resolution. For each hour, the best data point was selected using a simple ``majority rule'' algorithm. The scripts printed a new row every time at least one parameter had a new value in a given hour; for parameters for which no new value was read, the corresponding column was left blank.

The next step was to edit the files by hand, in order to remove corrupt data. Before editing, a backup copy of the raw data file was made, as a means of documenting exactly which, and how many, data points were removed. To facilitate the hand-editing process, the data files were fed to a plotting tool ({\tt gnuplot}), which made it easy to visually identify outliers. Any row that contained obviously corrupt data was removed from the file. By ``obviously corrupt'', we mean data points that contained physically meaningless values; e.g., a temperature of $-1000^\circ$F, a sudden spin-up to 15~rpm and back down to 5~rpm, etc. If there was any doubt as to whether a data point was corrupt or not, the interactive Windows application was used to examine the corresponding telemetry frames in their entirety; if the telemetry frame contained completely bogus data, the questionable data point was rejected, otherwise (if the telemetry frame appeared normal) the data point was kept. Very few data points were removed as a result of this process.

For some data sets (notably, spin and maneuvers) we also had information not based on telemetry. In these cases, the edited data files were imported into Microsoft Excel, and were reconciled with the alternate data source.

For some data sets, the physically meaningful quantity did not always come from the same sensor. For instance, the power of the transmitted radio beam came from either the TWT A or TWT B transmitted power reading depending on which TWT was operating. In these cases, the raw data files contained data for both parameters, regardless of their validity; however, in the edited data file, they were reconciled into a single column of data.

\subsection{Spin data}

The primary source for spin rate information on board Pioneer 10/11 was the spacecrafts' star sensor \cite{PC202}. This sensor generated a pulse every time a specific star (Canopus) came into its field of view. The time between pulses was measured by on-board electronics using an accuracy of 1/8192 seconds. This value was telemetered to the ground.

In addition to the star sensor, there were two separate sun sensors on board. By ground command, these could also be used for spin rate reference. Of the two sun sensors, only one had the ability to reliably detect a sun pulse when the angle between the Sun-spacecraft line and the spacecraft spin axes was small (down to 1$^\circ$.) The sun sensor was known to work reliably up to $\sim$33~AU.

The accuracy of the spin rate measurement was limited by the stability of the on-board clock (a 32,768 Hz crystal oscillator) and the phase error of the sensor instruments. An option, selectable by ground command, was to average the spin period measurements over several (typically 64) spacecraft revolutions. This was the preferred mode of operation for the spacecraft. In this mode, errors due to oscillator stability as well as phase errors are unlikely to be detectable in the telemetry (i.e., they add up to less than one half of 1/8192 seconds) and therefore, the telemetered spin rate can be considered accurate to within the resolution of the digital data.

The accuracy and the phase stability of the spin rate counter was not noticeably affected by the switch from the star sensor to the sun sensor or vice versa. Indeed, it is difficult to determine from telemetry which sensor was used at any given time; though certain telemetry bits supposedly provide this information, in actuality, the readings are ambiguous. Therefore, we do not distinguish between star sensor and sun sensor based spin rate measurements.

Spin rate information was telemetered to the ground using a set of 3 6-bit engineering words \cite{PC202}. Combined, these words gave the spin period is units of 1/8192 seconds:
\begin{equation}
t_\mathrm{SPIN}=\frac{C_\mathrm{405}}{2}+\frac{C_\mathrm{406}}{2^7}+\frac{C_\mathrm{407}}{2^{13}}
\label{eq:spin0}
\end{equation}

Engineering words were transmitted using one of two format types: if an accelerated engineering format was selected, several engineering words were transmitted simultaneously, whereas when a science format was chosen, only one engineering word per telemetry frame was transmitted in the form of a subcommutator.

For multi-word readings such as the spin rate, this means that unless all the words that constitute a reading are transmitted simultaneously, there was no way to ensure that words, transmitted separately, were obtained as part of the same measurement. For this reason, the spin rate words were marked as ``accelerated frames only'' in Pioneer project documentation.

Unfortunately, since accelerated engineering frames were rarely used, this means that proper spin rate measurements were telemetered to the ground only on a few selected occasions.

In practice, however, the spin rate being a (very) slowly changing value, it does not usually matter if words were obtained at the same time or not. Furthermore, spin rate words were not suppressed in the subcommutator: therefore, it is possible to reconstruct spin rate readings for the entire missions. Indeed, there is reason to believe that throughout the mission, the Pioneer project was relying on spin rate words telemetered in the subcommutator to obtain the correct spin rate of the spacecraft.

\subsubsection{Pioneer 10 spin data}

Spin data for Pioneer 10 came from several sources.

The star sensor on Pioneer 10 failed after Jupiter encounter. Thereafter, the primary spin reference was one of the spacecraft's two sun sensors. The exact time when the switch from the star sensor to the sun sensor took place is not known; spin rate telemetry appears reliable and accurate (jitter free) both before and after Jupiter encounter (11 days worth of telemetry during Jupiter encounter is missing from the telemetry files.)

The Pioneer 10 sun sensor became temporarily inoperative in November 1983 when a CONSCAN maneuver inadvertently reduced the Sun-spacecraft-spin axis angle to $\sim$1/4$^\circ$. The sun sensor was turned back on in July 1985; it was switched off permanently in May 1986, in order to avoid missed sun pulses (and incorrect results after the sensor averaged pulse periods over several spacecraft revolutions.)

When no sun sensor or star sensor was operating, the spacecraft continued to generate ``roll index'' pulses at the rate of the last spin rate measurement.

For Pioneer 10, a total of 98,286 hourly spin readings were extracted from the telemetry. As a result of manual editing, 227 rows were removed from this total, representing approximately 0.33\% of the data collected.

After removing data for the periods when no on-board spin reference was functioning, the remaining data set contains 57,376 rows.

After 1986, spot measurements of the Pioneer 10 spin rate were made using the Imaging Photo-Polarimeter (IPP) instrument \cite{AIAA870502}. In one operating mode, the IPP could be commanded to begin taking an image when a source of sufficient magnitude entered the instrument's field of view. The time from the last roll index pulse until the moment the instrument was triggered could be measured. Taking several such measurements at different times, the difference in time could serve as an indicator of the deviation of the spacecraft's actual spin rate relative to the ``frozen'' rate of the roll index pulse.

These measurements were made periodically until October 1993, when there was no sufficient power left on the spacecraft to continue operating the IPP instrument.

Pioneer 10 spin telemetry data, in units of RPM (revolutions per minute) is presented in the form of a tab-delimited text file, {\tt 23spin\_valid.dat}. This file contains valid data records: i.e., the set of 57,376 spin telemetry records from which bad records and records containing ``frozen'' spin rate readings have been removed. The three columns of the file contain the date ({\tt DOW Mon  D hh:mm:ss yyyy}), the spin rate in RPM, and the contents of telemetry word $C_{431}$ that tells us which sensor was used to obtain the spin rate reading. (From \cite{PC202} we learn that the last two bits of this telemetry word identify the star sensor or either of the two sun sensors as the roll rate reference in use. In actuality, the bits do not seem to be correlated with our knowledge of the spacecraft's state, and may not be trusted.)

A complete set of Pioneer 10 spin data combined from all sources is presented in the Microsoft Excel format data file {\tt 23spin.xls}. The file contains one worksheet with five columns. The first column is the ground received time of the hourly batch from which the spin rate was extracted. The second column (labeled ``Spin (RPM)'') contains the spin rate measurement combined from all three sources. The third, fourth, and fifth columns (labeled ``RollIDX'', ``IPP'', and ``Est'', respectively) contain the information obtained from the on-board roll index pulse reference (i.e., the star sensor or Sun sensors), the IPP instrument, and (after 1993) from ground-based estimates. Note the decreasing accuracy (fewer decimal digits) towards the end of the data set.

\subsubsection{Pioneer 11 spin data}

The star sensor on board Pioneer 11 remained functional throughout the spacecraft's mission, and provided reliable spin telemetry from launch until the last received signal.

The spin rate of Pioneer 11 was highly anomalous. Early in the mission, a thruster malfunction caused Pioneer 11 to spin up to approximately 7.6~rpm. The spin rate of Pioneer 11 continued to increase afterwards. Looking at the telemetry closely, one can observe that the spin rate actually {\it decreased} during quiescent periods between maneuvers (consistently with the behavior of Pioneer 10), but increased suddenly during each maneuver. This is interpreted as a result of a slight misalignment, relative to the spin axis, of one or more maneuvering thrusters. The spin rate increase was significant; at the time when the last spin rate telemetry was received, Pioneer 11 was already spinning at 8.37~rpm.

For Pioneer 11, a total of 90,536 hourly spin readings were collected. During manual editing, 47 records were removed, representing 0.05\% of the total.

Pioneer 11 spin data is presented in the form of a tab-delimited text data file, {\tt 24spin\_edited.dat}. The file contains the 90,489 valid telemetry readings. The columns of the file are the date ({\tt DOW Mon  D hh:mm:ss yyyy}), the spin rate in RPM, and the contents of telemetry word $C_{431}$. (See the previous section for interpreting this word.)

\subsection{Maneuver data}

Information about maneuvers that were executed by Pioneer 10/11 is scarce. Some handwritten notes exist that list maneuvers, but the reliability and completeness of this record is in question. Therefore, it is desirable to reconstruct the spacecrafts' maneuver history from telemetry files.

For each thruster, a sensor detected pulses, which were counted by the ACS. One 6-bit word per cluster containing the thruster pulse count was regularly telemetered to the Earth.

Secondary telemetry included thruster and thruster cluster assembly temperatures. However, these temperature readings are not a very reliable means to identify when thrusters were fired. A few short pulses changed thruster temperatures very little, and the thrusters returned to equilibrium temperatures shortly afterwards. At the lowest telemetry rate, each telemetry word was transmitted to the Earth at 51.2 minute intervals, which means that many thruster events are not at all visible in thruster temperature telemetry.

For this reason, only the thruster pulse count should be accepted as reliable telemetry about thruster events. The extent to which this reading can be trusted is limited by several factors:
\begin{itemize}
\item There is evidence that, especially late in the mission, the sensor mechanism missed thruster pulse events. The evidence comes in the form of maneuver events that involved a pair of thrusters, only one of which registered pulses, yet the maneuver itself completed successfully, suggesting that both thrusters fired as planned. The missed events are attributed to the age of the thruster and sensor mechanism and the extreme cold environment in which the thrusters operated.
\item The thruster pulse count gives no indication as to when thruster events took place. The change in thruster pulse count relative to the previous reading provides a time window (in ground received time) when the events occurred. If this time window is large (e.g., due to missing telemetry records), several thruster events might be ``lumped'' together into a single increment in the pulse count.
\item As the thruster pulse counter is a 6-bit counter, the possibility exists that thruster events amounting to exactly $64n$ pulses ($n$ integer) are missed altogether. In practice, such occurrences would be rare if they exist at all, as most maneuvers utilized only a few pulses per thruster, certainly much less than 64; the time between maneuver events was sufficiently large for the new pulse count to be telemetered to the Earth first.
\end{itemize}

\subsubsection{Editing Strategy}

Thruster pulse counts were obtained from the raw telemetry files using one-hour batches. This editing strategy, which we used for all telemetry, significantly reduced noise due to corrupt data, without introducing a significant loss in time resolution. For each one-hour batch, readings of a specific telemetry word were collected, however many times that telemetry word occurred in the records. If more than one reading was found, the value that occurred most often (majority rule) was taken to be the valid reading for that hour.

For thruster pulse counts, only {\it changes} in the reading matter; a constant pulse count value indicates no thruster event and does not need to be reported. Therefore, the script program that was used to extract thruster data generated one output line every time a changed pulse count value was detected for any of the four VPTs.

The resulting file was short (several hundred lines only, for each spacecraft) and almost free of errors. The few errors (corrupt data) that remain were easily identifiable, as they involved consecutive pulse count readings that added up to $0_{\mathrm{mod}64}$ for each thruster. As an example, consider the following lines from the telemetry:
\begin{small}
\begin{verbatim}
 Sat Aug 15 23:00:00 1998        Sun Aug 16 00:00:00 1998        18      57      35      7
 Sun Aug 16 00:00:00 1998        Sun Aug 16 02:00:00 1998        46      7       29      57
\end{verbatim}
\end{small}

What these lines indicate is that there was a set of bogus pulse count values at or around 11 PM on August 15, 1998; however, in the next hourly batch, the original (correct) pulse count values were reported. No maneuver has taken place. Sets of lines such as this example were filtered from the telemetry.

A further editing step results from the fact that pulse counts for individual thrusters were not telemetered in the same telemetry frame; therefore, for the same maneuver, it is possible that pulse counts for participating thrusters arrived a significant time apart, and may have appeared in different hourly batches. Consecutive output lines that obviously represented the same maneuver were reconciled into a single output line, with the start/stop timestamps appropriately adjusted. As an example, the following pair of lines
\begin{small}
\begin{verbatim}
 Sun Mar 28 13:00:00 1976        Sun Mar 28 14:00:00 1976        0       11      0       0
 Sun Mar 28 14:00:00 1976        Sun Mar 28 15:00:00 1976        0       0       0       11
\end{verbatim}
\end{small}
was reconciled into
\begin{small}
\begin{verbatim}
 Sun Mar 28 13:00:00 1976        Sun Mar 28 15:00:00 1976        0       11      0       11
\end{verbatim}
\end{small}

\subsubsection{Reconciliation with Other Data Sources}

Where information not based on telemetry was available on maneuvers, we were able to successfully reconcile these records with the telemetry information. The notes, etc. containing maneuver information typically contained the type of the maneuver, and precise time of the first and last pulse fired, and the number of pulses fired. In most instances, we were able to match this data against telemetry records. However, the telemetry also contained information about maneuvers not appearing in these notes. The nature of the telemetry record is such that we have reason to believe that these maneuvers did, in fact, take place; a spurious reading in the telemetry would have arrived in the form of a ``one off'' value, with the pulse count returning to the original value in subsequent batches, resulting in consecutive readings that add up to $0_{\mathrm{mod}64}$.

\subsubsection{Maneuver files}

After data extraction and the manual editing has been completed, a total of 377 propulsion events were identifiable in the telemetry for Pioneer 10, and 477 propulsion events for Pioneer 11. The majority of these were precession maneuvers in the ``positive'' direction.

We have available limited propulsion maneuver data from other sources, covering the time period between 1987 and 1998 for Pioneer 10, 1987-1990 for Pioneer 11, and some additional data points from 1981. For most maneuvers, there is good agreement between the two data sets, though there are several discrepancies.

The reconciled set of maneuver data is stored in the Microsoft Excel format files {\tt 23pulse.xls} and {\tt 24pulse.xls}. In addition to the columns present in the tab-delimited text file, for extra columns are presented in each file: day-of-the-year, number of pulses, first pulse time stamp, and last pulse time stamp for recorded maneuvers. (Not all of this information is available for all recorded maneuvers.) Where the day-of-the-year differs from the day-of-the-year of the corresponding telemetry record, it is marked typographically (bold of the day-of-the-year value is greater than the telemetry date, bold and red if it is earlier). Extra rows are inserted where the alternate data source indicates a maneuver that cannot be seen in telemetry.

\subsection{Temperature data}

Of the large number of temperature sensors that are present on board Pioneer 10/11, we selected 10 readings that we believe provide readings not specific to the interior of a component or science instrument, but more representative of the overall temperature profile of the spacecraft. Four of these readings are RTG fin root temperature sensor readings, and provide information on the temporal evolution of RTG thermal profiles. The remaining six are platform temperature sensors situated at various locations inside the spacecraft body, providing readings on the temperature of the electronics platform therein.

The total number of hourly temperature records retrieved for Pioneer 10 is 95,882. Only three of these records were removed by hand for containing obviously bad data. The resulting data set of 95,879 records is stored in the file {\tt 23temp\_edited.dat}.

For Pioneer 11, the total number of hourly temperature records is 88,833. Only one record was removed as containing bogus data, leaving 88,832 records in the resulting file, {\tt 24temp\_edited.dat}.

\subsection{Electrical subsystem data}

The electrical subsystem of the Pioneer 10/11 spacecraft is complex, and several electrical readings are available in the telemetry. The information that is required to evaluate the electrical system from a thermal perspective is the total amount of electrical power on board, and the amount of electrical power turned into waste heat in the main spacecraft compartment.

The total electrical power on board can be computed from electrical readings that are available for each RTG. There exists a voltage and a current reading for each generator. Summing them up, we can compute the total amount of power:
\begin{equation}
P_\mathrm{RTG}=\sum_{i=1}^4{U_iI_i}.
\end{equation}

Not all of this electrical power reaches the spacecraft's power supply circuitry. The RTGs are connected to the main spacecraft body by long ribbon cable that has a known electrical resistance. Knowing the RTG current, it is possible to calculate the voltage drop and the corresponding loss of electrical power on these cables\footnote{According to \cite{PC202}, the RTG voltage is measured at the input terminals of the inverter circuits, which means that these values are already reduced by the voltage drop along the ribbon cable. Actual examination of the telemetry values does suggest, however, that the opposite may be true; namely, that the RTG voltage telemetry has been calibrated to account for the cable loss, and the telemetered values represent the voltages that would be measured at the RTG terminals.} The electrical resistance along the cable to the near and far RTGs is $R_\mathrm{inboard}=0.017$ and $R_\mathrm{outboard}=0.021\,\mathrm{\Omega}$, respectively. The cable loss can, therefore, be estimated as
\begin{equation}
P_\mathrm{cable}=R_\mathrm{inboard}\sum_{i=3,4}I_i+R_\mathrm{outboard}\sum_{i=1,2}I_i.
\end{equation}

Most electrical power is turned into waste heat inside the spacecraft body. There are, however, a few exceptions. First, the shunt circuit's radiating plate, which is mounted on the exterior of the spacecraft body, off to one side. The shunt radiator is characterized in \cite{PC202}, and from the diagrams it is apparent that it is a $R_\mathrm{shunt}=5.25\,\mathrm{\Omega}$ resistor. The shunt current is known from telemetry. The total shunted power can be computed using this value and the main bus voltage; the power radiated by the radiating plate can be computed using the shunt current and $R_\mathrm{shunt}$:
\begin{eqnarray}
P_\mathrm{shunt}&=&I_\mathrm{shunt}U_\mathrm{bus}\\
P_\mathrm{shrad}&=&I_\mathrm{shunt}^2R_\mathrm{shunt}^{}.
\end{eqnarray}

As the main bus voltage and main bus current are known from telemetry, we can also calculate the main bus power:
\begin{equation}
P_\mathrm{bus}=U_\mathrm{bus}I_\mathrm{bus}.
\end{equation}

We can also calculate battery power. The battery voltage is telemetered, as is the battery charge current. A telemetry bit tells us whether or not the battery is being discharged; the discharge current is also telemetered:
\begin{equation}
P_\mathrm{bat}=U_\mathrm{bat}\left(I_\mathrm{charge}-I_\mathrm{discharge}\right),
\end{equation}
where $I_\mathrm{discharge}$ is set to zero (i.e., the actual telemetered value is ignored) if battery discharge is not enabled.

There are several science instruments mounted outside the main spacecraft body. If one wishes to calculate the total amount of heat dissipated by the spacecraft body, it is necessary to subtract the power consumption of externally mounted instruments from the electrical power that enters the spacecraft body\footnote{Important: the radio transmitter power must also be subtracted.}.

\begin{table}
\caption{Computed and raw telemetry values in the electrical telemetry files.\label{tb:eleccols}}
\centering
\begin{tabular}{|cc|}\hline
Field&Description\\\hline\hline
Date&Ground received time\\
$P_\mathrm{RTG}$&Total RTG electrical power\\
$P_\mathrm{cable}$&Power loss along RTG ribbon cable\\
$U_\mathrm{bus}$&Main bus voltage\\
$P_\mathrm{bus}$&Main bus power\\
$P_\mathrm{shunt}$&Shunt power\\
$P_\mathrm{shrad}$&Shunt radiator plate power\\
$P_\mathrm{bat}$&Battery power\\
$\mathrm{State}_{108}$&Instrument power bits from $C_{108}$\\
$\mathrm{State}_{124}$&Instrument power bits from $C_{124}$\\
$\mathrm{State}_{128}$&Subsystem power bits (bits 1--3) from $C_{128}$\\
$\mathrm{State}_{316}$&Subsystem power bits (bits 1--2) from $C_{316}$\\
$\mathrm{State}_{421}$&Subsystem power bits (bit 4) from $C_{421}$\\\hline
$|$&Separator character\\\hline
$C_{105}$--$C_{110}$&...\\
$C_{113}$--$C_{114}$&\\
$C_{122}$--$C_{129}$&Raw telemetry words\\
$C_{131}$&\\
$C_{316}$&\\
$C_{421}$&...\\
\hline
\end{tabular}
\end{table}

To address these complexities of the spacecraft's electrical subsystem, the electrical telemetry data file that we developed contains many columns. The first several columns contain calculated values; however, the original telemetry values are also included in additional columns, in order to ensure that the calculations can be reproduced. In sum, the electrical telemetry file contains the columns specified in Table~\ref{tb:eleccols}.

The total number of hourly electrical telemetry records that we obtained for Pioneer 10 is 96,800. During the editing phase, we removed 147 records because they contained obviously bogus data, leaving a total of 96,653 records in the edited telemetry file, {\tt 23elec\_edited.dat}.

For Pioneer 11, we obtained 89,350 hourly electrical telemetry records, from which 26 records were removed manually. The result is 89,324 records in the edited electrical telemetry file, {\tt 24elec\_edited.dat}.

\subsection{Communication subsystem data}

The spacecrafts' radio transmitter and receiver assemblies have many telemetered parameters, but only a few that are of interest to us.

The most important is the transmitter power, which was measured on board and telemetered in the form of a value measured in dBm. This can be correlated with the TWT temperature and cathode current values to verify that the readings are consistent.

The characteristic parameter of the spacecraft's receiver is the received signal strength, also measured in units of dBm.

For Pioneer 10, TWT A and receiver A were used throughout the mission. For this reason, we retrieved values for this transmitter and receiver only.

For Pioneer 11, the transmitter was switched from TWT A to TWT B on May 13, 1973, and TWT B remained the operating transmitter through the rest of the mission. The receiver was also switched from receiver A to receiver B on January 27, 1984; subsequently, receiver B was used for the remainder of the mission.

\begin{table}
\caption{Communications subsystem telemetry.\label{tb:commtel}}
\vskip 6pt
\centering
\begin{tabular}{|ccc|}\hline
Parameter&A&B\\\hline\hline
TWT temperature&$C_{205}$&$C_{228}$\\
TWT cathode current&$C_{208}$&$C_{215}$\\
TWT power&$C_{231}$&$C_{214}$\\
Receiver power&$C_{232}$&$C_{213}$\\\hline
\end{tabular}
\end{table}

For this reason, we retrieve parameters for both transmitters and both receivers for Pioneer 11.

The parameters retrieved for the communication subsystem are summarized in Table~\ref{tb:commtel}. For Pioneer 10, a total of 93,274 hourly telemetry records were retrieved, of which 113 were removed manually as they contained obviously bogus data, leaving 93,161 records in the edited communication subsystem telemetry file, {\tt 23comm\_edited.dat}. For Pioneer 11, a total of 88,229 records were retrieved, of which 564 records were removed because they either contained bogus data, or (after the TWT A/TWT B, or receiver A/receiver B data were reconciled) no data, leaving 87,665 records in the edited communication subsystem telemetry file {\tt 24comm\_edited.dat}.

In addition to the communications subsystem data from on board the spacecraft, MDRs also contain valuable information produced at the DSN receiving station. Most notably, this includes the DSN AGC that measures the amount of amplification applied at the DSN to normalize the received signal level, and thus provides an indirect measure of the received signal strength.

AGC is measured in dB, relative to a nominal signal strength of 0~dBm at which no amplification occurs. In other words, the received signal strength would be $-1$ times the AGC value, measured in dBm. The AGC value is encoded as a 32-bit fixed-point binary value, with 4 binary digits right of the binary point, resulting in a resolution of 1/16~Hz.

For both Pioneer 10 and 11, well over 100 million AGC readings are present in the raw data set. A more manageable data set was produced by calculating hourly averages with outliers (defined as AGC values falling outside the range of 1-255~dB) removed. These results are stored in the files {\tt 23AGC.dat} and {\tt 24AGC.dat}.

The DSN AGC is a function of several factors, some of which are known. Specifically, the distance to the spacecraft is known, as is the transmitted power (telemetry words $C_{231}$ and $C_{214}$). An additional factor is the collecting area of the DSN antenna that was used to receive the signal; this, too, is known as the DSN station identifier is recorded along with the AGC value, and the antenna diameters are known even for historic DSN stations. It is, therefore, possible to normalize the AGC signal by compensating for these three factors. The expression
\begin{equation}
P_\mathrm{AGC}'=P_\mathrm{AGC}-20\log_{10} R+P_\mathrm{TWT}+20\log_{10}(D/70),
\end{equation}
where $P_\mathrm{AGC}$ is the measured AGC value, $R$ is the geocentric distance to the spacecraft, and $D$ is the DSN antenna diameter, normalizes the signal relative to a 1~mW transmitter at 1~AU from the Earth, as received by a 70~m antenna. The normalized values are available in the files {\tt 23AGCcomm.dat} and {\tt 24AGCcomm.dat}.

\subsection{Propulsion subsystem data}

In addition to the pulse count/maneuver data discussed earlier, the propulsion subsystem can be characterized in other ways that may be useful for the investigation of the Pioneer anomaly.

First, in order to corroborate readings obtained from pulse counts or other sources, it may be necessary to examine velocity and precession thruster and thruster cluster temperatures. A total of six readings (four thruster temperatures, two thruster cluster temperatures) are available. These readings were collected for both spacecraft. For Pioneer 10, 93,602 hourly thruster temperature records were collected, of which 25 were removed for bad data, leaving 93,577 records in the edited thruster temperature file {\tt 23ttemp\_edited.dat}. For Pioneer 11, the total number of thruster temperature hourly records was 87,949; only one record was removed for bad data. The resulting file, containing 87,948 records, is called {\tt 24ttemp\_edited.dat}.

Additional parameters that characterize the spacecraft's propulsion system are the temperature of the N$_2$ pressurant in the propellant tank; the temperature and pressure of the hydrazine propellant; and the temperature of the spin thruster cluster assembly. These readings were collected in a separate file for each spacecraft. For Pioneer 10, a total of 95,828 hourly records were collected for the propulsion system, of which 57 were removed for containing bad data, leaving 95,771 records in the file {\tt 23prop\_edited.dat.} For Pioneer 11, the total number of hourly propulsion system records was 89,029; after removing 11 bad records, we were left with 89,018 records in {\tt 24prop\_edited.dat}.

\subsection{Data files summary}

\begin{table}[ht!]
\caption{\normalsize Telemetry data files and explanatory notes.\label{tb:files}}
\small \centering
\begin{tabular}{|ll|}\hline
File name&Description\\\hline\hline
\multicolumn{2}{|l|}{EXPLANATORY FILES}\\\hline
{\tt index.html}&CD/DVD index in HTML format\\
{\tt pioneer\_data.pdf}&Explanatory Supplement\\
{\tt pioneer\_spinnote.pdf}&Spin rate notes\\
{\tt pioneer\_pulsenote.pdf}&Thruster pulse count notes\\
{\tt pioneer\_agcnote.pdf}&AGC notes\\\hline\hline
\multicolumn{2}{|l|}{SPIN DATA}\\\hline
{\tt DATA$\backslash$23spin\_valid.dat}&Pioneer 10 spin rate (RPM) hourly readings\\
{\tt DATA$\backslash$23spin.xls}&Pioneer 10 reconciled spin data (Excel format)\\
{\tt DATA$\backslash$24spin\_edited.dat}&Pioneer 11 spin rate (RPM) hourly readings\\\hline\hline
\multicolumn{2}{|l|}{THRUSTER PULSE COUNTS}\\\hline
{\tt DATA$\backslash$23pulse\_edited.dat}&Pioneer 10 thruster pulse counts\\
{\tt DATA$\backslash$24pulse\_edited.dat}&Pioneer 11 thruster pulse counts\\
{\tt DATA$\backslash$23pulse.xls}&Pioneer 10 reconciled counts (Excel format)\\
{\tt DATA$\backslash$24pulse.xls}&Pioneer 11 reconciled counts (Excel format)\\\hline\hline
\multicolumn{2}{|l|}{THRUSTER TEMPERATURES}\\\hline
{\tt DATA$\backslash$23ttemp\_edited.dat}&Pioneer 10 thruster temperatures\\
{\tt DATA$\backslash$24ttemp\_edited.dat}&Pioneer 11 thruster temperatures\\\hline\hline
\multicolumn{2}{|l|}{ELECTRICAL READINGS}\\\hline
{\tt DATA$\backslash$23elec\_edited.dat}&Pioneer 10 electrical readings\\
{\tt DATA$\backslash$24elec\_edited.dat}&Pioneer 11 electrical readings\\\hline\hline
\multicolumn{2}{|l|}{COMMUNICATIONS SUBSYSTEM}\\\hline
{\tt DATA$\backslash$23comm\_edited.dat}&Pioneer 10 communications subsystem readings\\
{\tt DATA$\backslash$24comm\_edited.dat}&Pioneer 11 communications subsystem readings\\
{\tt DATA$\backslash$24comm.dat}&Pioneer 11 raw communication data\\\hline\hline
\multicolumn{2}{|l|}{AGC}\\\hline
{\tt DATA$\backslash$23AGC.dat}&Pioneer 10 DSN AGC\\
{\tt DATA$\backslash$24AGC.dat}&Pioneer 11 DSN AGC\\
{\tt DATA$\backslash$23AGCcomm.dat}&Pioneer 10 DSN AGC (normalized)\\
{\tt DATA$\backslash$24AGCcomm.dat}&Pioneer 11 DSN AGC (normalized)\\\hline\hline
\multicolumn{2}{|l|}{PROPULSION}\\\hline
{\tt DATA$\backslash$23prop\_edited.dat}&Pioneer 10 miscellaneous propulsion readings\\
{\tt DATA$\backslash$24prop\_edited.dat}&Pioneer 11 miscellaneous propulsion readings\\\hline\hline
\multicolumn{2}{|l|}{PLATFORM TEMPERATURES}\\\hline
{\tt DATA$\backslash$23temp\_edited.dat}&Pioneer 10 RTG and platform temperatures\\
{\tt DATA$\backslash$24temp\_edited.dat}&Pioneer 11 RTG and platform temperatures\\\hline
\end{tabular}
\end{table}

Telemetry data files released on CD/DVD are summarized in Table~\ref{tb:files}.

\appendix

\section*{Appendices}

\section{Pioneer 10/11 Maneuver Data (thruster pulse counts)}


\subsection{Spacecraft Maneuvers}

The Pioneer 10/11 spacecraft had 6 thrusters each \cite{PC202}, mounted along the rim of the high gain antenna (HGA) in three thruster cluster assemblies. One pair of thrusters was mounted in the antenna plane, providing spin/despin control. The remaining two pairs were mounted on opposite sides of the HGA, perpendicular to the antenna plane, providing velocity and precession control.

Thruster valves were operated by the spacecrafts' attitude control subsystem (ACS). Thrusters were fired in the form of pulses of varying duration.

The Pioneer 10/11 spacecraft were designed to perform three types of maneuvers:
\begin{itemize}
\item{\it Spin/despin maneuvers} were used, at the beginning of mission, to alter the spacecraft's spin rate. One or the other thruster of the spin/despin thruster cluster assembly was used to effect a spin-up or spin-down of the spacecraft.
\item{\it Delta-v ($\mathrm{\Delta v}$) maneuvers} were used to change the spacecraft's trajectory. With the spacecraft pointing in the right direction, two velocity and precession thrusters (VPTs) on opposite sides of the high gain antenna (HGA), pointing in the same direction, were fired, accelerating the spacecraft along the spin axis.
\item{\it Precession maneuvers} were used to change the orientation of the spacecraft's spin axis. Two velocity and precession thrusters, mounted on opposite sides of the HGA and firing in opposing directions at a specific roll phase of the spacecraft could be used to precisely alter the orientation of the spacecraft spin axis.
\end{itemize}

Thruster firings could take place under ground control, or as part of the closed-loop ``CONSCAN'' maneuver that provided a means for the spacecraft to automatically ``home in'' on the DSN carrier signal received from the Earth. In all cases, every time a thruster was fired, the corresponding thruster pulse count register was incremented by one.

VPTs were numbered 1 through 4, or alternatively, labeled as follows:

\begin{center}
\begin{tabular}{c@{: }c}
VPT 1&1A\\
VPT 2&1B\\
VPT 3&2B\\
VPT 4&2A\\
\end{tabular}
\end{center}

VPTs were used in the following combinations:

\begin{center}
\begin{tabular}{r@{: }c}
``Positive'' precession&2 and 3 (1B and 2B)\\
``Negative'' precession&1 and 4 (1A and 2A)\\
Fore $\mathrm{\Delta}$v&1 and 3 (1A and 2B)\\
Aft $\mathrm{\Delta}$v&2 and 4 (1B and 2A)\\
\end{tabular}
\end{center}

Information about maneuvers that were executed by Pioneer 10/11 is scarce. Some handwritten notes exist that list maneuvers, but the reliability and completeness of this record is in question. Therefore, it is desirable to reconstruct the spacecrafts' maneuver history from telemetry files.

\subsection{Thruster Telemetry}

For each thruster, a sensor detected pulses, which were counted by the ACS. One 6-bit word per cluster containing the thruster pulse count was regularly telemetered to the Earth.

Secondary telemetry included thruster and thruster cluster assembly temperatures. However, these temperature readings are not a very reliable means to identify when thrusters were fired. A few short pulses changed thruster temperatures very little, and the thrusters returned to equilibrium temperatures shortly afterwards. At the lowest telemetry rate, each telemetry word was transmitted to the Earth at 51.2 minute intervals, which means that many thruster events are not at all visible in thruster temperature telemetry.

Similarly, indirect indicators of thruster firings (e.g., change in propellant tank pressure or temperature) cannot be relied upon; for maneuvers that involved only a few thruster pulses, there were no detectable changes in propellant tank readings.

For this reason, only the thruster pulse count should be accepted as reliable telemetry about thruster events. The extent to which this reading can be trusted is limited by several factors:
\begin{itemize}
\item There is evidence that, especially late in the mission, the sensor mechanism missed thruster pulse events. The evidence comes in the form of maneuver events that involved a pair of thrusters, only one of which registered pulses, yet the maneuver itself completed successfully, suggesting that both thrusters fired as planned. The missed events are attributed to the age of the thruster and sensor mechanism and the extreme cold environment in which the thrusters operated.
\item The thruster pulse count gives no indication as to when thruster events took place. The change in thruster pulse count relative to the previous reading provides a time window (in ground received time) when the events occurred. If this time window is large (e.g., due to missing telemetry records), several thruster events might be ``lumped'' together into a single increment in the pulse count.
\item As the thruster pulse counter is a 6-bit counter, the possibility exists that thruster events amounting to exactly $64n$ pulses ($n$ integer) are missed altogether. In practice, such occurrences would be rare if they exist at all, as most maneuvers utilized only a few pulses per thruster, certainly much less than 64; the time between maneuver events was sufficiently large for the new pulse count to be telemetered to the Earth first.
\end{itemize}

\subsection{Editing Strategy}

The limitations in thruster telemetry also suggested a straightforward editing strategy to reduce the raw telemetry data files to a maneuver list of manageable size.

Thruster pulse counts were obtained from the raw telemetry files using one-hour batches. This editing strategy, which we used for all telemetry, significantly reduced noise due to corrupt data, without introducing a significant loss in time resolution. For each one-hour batch, readings of a specific telemetry word were collected, however many times that telemetry word occurred in the records. If more than one reading was found, the value that occurred most often (majority rule) was taken to be the valid reading for that hour.

For thruster pulse counts, only {\it changes} in the reading matter; a constant pulse count value indicates no thruster event and does not need to be reported. Therefore, the script program that was used to extract thruster data generated one output line every time a changed pulse count value was detected for any of the four VPTs.

The resulting file was short (several hundred lines only, for each spacecraft) and almost free of errors. The few errors (corrupt data) that remain were easily identifiable, as they involved consecutive pulse count readings that added up to $0_{\mathrm{mod}64}$ for each thruster. As an example, consider the following lines from the telemetry:
\begin{small}
\begin{verbatim}
 Sat Aug 15 23:00:00 1998        Sun Aug 16 00:00:00 1998        18      57      35      7
 Sun Aug 16 00:00:00 1998        Sun Aug 16 02:00:00 1998        46      7       29      57
\end{verbatim}
\end{small}

What these lines indicate is that there was a set of bogus pulse count values at or around 11 PM on August 15, 1998; however, in the next hourly batch, the original (correct) pulse count values were reported. No maneuver has taken place. Sets of lines such as this example were filtered from the telemetry.

A further editing step results from the fact that pulse counts for individual thrusters were not telemetered in the same telemetry frame; therefore, for the same maneuver, it is possible that pulse counts for participating thrusters arrived a significant time apart, and may have appeared in different hourly batches. Consecutive output lines that obviously represented the same maneuver were reconciled into a single output line, with the start/stop timestamps appropriately adjusted. As an example, the following pair of lines
\begin{small}
\begin{verbatim}
 Sun Mar 28 13:00:00 1976        Sun Mar 28 14:00:00 1976        0       11      0       0
 Sun Mar 28 14:00:00 1976        Sun Mar 28 15:00:00 1976        0       0       0       11
\end{verbatim}
\end{small}
was reconciled into
\begin{small}
\begin{verbatim}
 Sun Mar 28 13:00:00 1976        Sun Mar 28 15:00:00 1976        0       11      0       11
\end{verbatim}
\end{small}

\subsection{Reconciliation with Other Data Sources}

Where information not based on telemetry was available on maneuvers, we were able to successfully reconcile these records with the telemetry information. The notes, etc. containing maneuver information typically contained the type of the maneuver, and precise time of the first and last pulse fired, and the number of pulses fired. In most instances, we were able to match this data against telemetry records. However, the telemetry also contained information about maneuvers not appearing in these notes. The nature of the telemetry record is such that we have reason to believe that these maneuvers did, in fact, take place; a spurious reading in the telemetry would have arrived in the form of a ``one off'' value, with the pulse count returning to the original value in subsequent batches, resulting in consecutive readings that add up to $0_{\mathrm{mod}64}$.

\subsection{Pioneer 10 Maneuvers}

\begin{table}
\caption{Annual number of propulsive maneuver events from telemetry, for Pioneer 10/11.}
\label{tb:annual}
\centering
\vskip 12pt
\begin{tabular}{|l|c|c|}\hline
Year&Pioneer 10&Pioneer 11\\\hline\hline
1972&89&\\
1973&76&69\\
1974&46&88\\
1975&44&85\\
1976&15&54\\
1977&10&27\\
1978&9&23\\
1979&11&24\\
1980&9&13\\
1981&6&16\\
1982&7&7\\
1983&7&9\\
1984&7&10\\
1985&9&7\\
1986&5&6\\
1987&4&6\\
1988&4&7\\
1989&3&8\\
1990&2&5\\
1991&1&4\\
1992&1&4\\
1993&4&3\\
1994&2&2\\
1995&2&\\
1996&2&\\
1997&1&\\
1998&1&\\\hline\hline
TOTAL:&377&477\\\hline
\end{tabular}
\end{table}

After data extraction and the manual editing has been completed, a total of 377 propulsion events were identifiable in the telemetry (Table~\ref{tb:annual}). The majority of these were precession maneuvers in the ``positive'' direction.

We have available limited propulsion maneuver data from other sources, covering the time period between 1987 and 1998, and some additional data points from 1981. For most maneuvers, there is good agreement between the two data sets, though there are several discrepancies.

Many of these discrepancies may be due to transcription errors, as maneuver data was recorded on handwritten slips. By way of example, the telemetry shows that a maneuver took place on or about July 7, 1987; the alternate data set shows a maneuver on June 7, 1987.

Some discrepancies very late in the mission may be due to sensor errors. We have reason to believe that in some cases, due to the extreme cold temperatures, the sensor transducers failed to register a thruster pulse event. One maneuver in 1994, one in 1997, and one in 1998 is recorded in the alternate data set, but cannot be seen in the telemetry. It is quite possible that the maneuvers did, in fact, take place and the thruster pulse counters remained unchanged due to this sensor malfunction.

There are also discrepancies in time stamps. This may, in some cases, be due to the fact that it could take many hours before a pulse count change is telemetered to the ground after a thruster firing event. However, in some cases the telemetry pulse count change actually precedes the date of the maneuver recorded in the alternate data set. Since it is highly unlikely that a malfunction caused a pulse count to register before the actual maneuver event, a more likely explanation is that the manual recording of maneuvers was erroneous.

The filtered and edited set of Pioneer 10 pulse count telemetry is recorded in a tab-delimited text file named {\tt 23pulse\_edited.dat}. The file has six columns: a start and stop date that bracket the hourly batch(es) during which the pulse count change occurred, and pulse count changes for thrusters 1A, 1B, 2A, 2B, in this order.

The reconciled set of maneuver data is stored in the Microsoft Excel format file {\tt 23pulse.xls}. In addition to the columns present in the tab-delimited text file, for extra columns are presented: day-of-the-year, number of pulses, first pulse time stamp, and last pulse time stamp for recorded maneuvers. (Not all of this information is available for all recorded maneuvers.) Where the day-of-the-year differs from the day-of-the-year of the corresponding telemetry record, it is marked typographically (bold of the day-of-the-year value is greater than the telemetry date, bold and red if it is earlier). Extra rows are inserted where the alternate data source indicates a maneuver that cannot be seen in telemetry.

\subsection{Pioneer 11 Maneuvers}

After data extraction and the manual editing has been completed, a total of 477 propulsion events were identifiable in the telemetry (Table~\ref{tb:annual}). The majority of these were precession maneuvers in the ``positive'' direction.

As for Pioneer 10, we also have limited propulsion maneuver data for Pioneer 11 collected from other sources, covering the time period between 1987 and 1990, and some additional data points from 1981. The agreement between this information and the telemetry is very good. All recorded maneuvers are visible in the telemetry data. Even date/time stamps appear to be in good agreement.

The filtered and edited set of Pioneer 11 pulse count telemetry is recorded in a tab-delimited text file named {\tt 24pulse\_edited.dat}. The file has six columns: a start and stop date that bracket the batch during which the pulse count change occurred, and pulse count changes for thrusters 1A, 1B, 2A, 2B, in this order.

The reconciled set of maneuver data is stored in the Microsoft Excel format file {\tt 24pulse.xls}. In addition to the columns present in the tab-delimited text file, for extra columns are presented: day-of-the-year, number of pulses, first pulse time stamp, and last pulse time stamp for recorded maneuvers. (Not all of this information is available for all recorded maneuvers.) Where the day-of-the-year differs from the day-of-the-year of the corresponding telemetry record, it is marked typographically (bold of the day-of-the-year value is greater than the telemetry date, bold and red if it is earlier). Extra rows are inserted where the alternate data source indicates a maneuver that cannot be seen in telemetry.


\section{Pioneer 10/11 Spin Data}

\subsection{Spin Telemetry}

The primary source for spin rate information on board Pioneer 10/11 was the spacecrafts' star sensor \cite{PC202}. This sensor generated a pulse every time a specific star (Canopus) came into its field of view. The time between pulses was measured by on-board electronics using an accuracy of 1/8192 seconds. This value was telemetered to the ground.

In addition to the star sensor, there were two separate sun sensors on board. By ground command, these could also be used for spin rate reference. Of the two sun sensors, only one had the ability to reliably detect a sun pulse when the angle between the Sun-spacecraft line and the spacecraft spin axes was small (down to 1$^\circ$.) The sun sensor was known to work reliably up to $\sim$33~AU.

The accuracy of the spin rate measurement was limited by the stability of the on-board clock (a 32,768 Hz crystal oscillator) and the phase error of the sensor instruments. An option, selectable by ground command, was to average the spin period measurements over several (typically 64) spacecraft revolutions. This was the preferred mode of operation for the spacecraft. In this mode, errors due to oscillator stability as well as phase errors are unlikely to be detectable in the telemetry (i.e., they add up to less than one half of 1/8192 seconds) and therefore, the telemetered spin rate can be considered accurate to within the resolution of the digital data.

The accuracy and the phase stability of the spin rate counter was not noticeably affected by the switch from the star sensor to the sun sensor or vice versa. Indeed, it is difficult to determine from telemetry which sensor was used at any given time; though certain telemetry bits supposedly provide this information, in actuality, the readings are ambiguous. Therefore, we do not distinguish between star sensor and sun sensor based spin rate measurements.

Spin rate information was telemetered to the ground using a set of 3 6-bit engineering words \cite{PC202}. Combined, these words gave the spin period is units of 1/8192 seconds:
\begin{equation}
t_\mathrm{SPIN}=\frac{C_\mathrm{405}}{2}+\frac{C_\mathrm{406}}{2^7}+\frac{C_\mathrm{407}}{2^{13}}
\label{eq:spin}
\end{equation}

Engineering words were transmitted using one of two format types: if an accelerated engineering format was selected, several engineering words were transmitted simultaneously, whereas when a science format was chosen, only one engineering word per telemetry frame was transmitted in the form of a subcommutator.

For multi-word readings such as the spin rate, this means that unless all the words that constitute a reading are transmitted simultaneously, there was no way to ensure that words, transmitted separately, were obtained as part of the same measurement. For this reason, the spin rate words were marked as ``accelerated frames only'' in Pioneer project documentation.

Unfortunately, since accelerated engineering frames were rarely used, this means that proper spin rate measurements were telemetered to the ground only on a few selected occasions.

In practice, however, the spin rate being a (very) slowly changing value, it does not usually matter if words were obtained at the same time or not. Furthermore, spin rate words were not suppressed in the subcommutator: therefore, it is possible to reconstruct spin rate readings for the entire missions. Indeed, there is reason to believe that throughout the mission, the Pioneer project was relying on spin rate words telemetered in the subcommutator to obtain the correct spin rate of the spacecraft.

The algorithm we use to obtain a spin rate measurement can be summarized as follows. We divide the raw telemetry into 1-hour batches, and collect values for all three spin rate words. If all three appeared at least once, we assume that we have a valid spin reading at that hour. If a word was telemetered more than once, we take a ``majority rule'' approach and use the value that appeared most often during the hour. Then, we use the three spin rate words to construct the 18-bit spin rate reading as defined in Eq. \ref{eq:spin}.

There are two sources of errors that can result in a corrupt spin rate reading. First, the raw telemetry may contain errors; especially late in the mission, there are many corrupt raw telemetry frames. Second, if a spin rate change occurred that caused a carry-over across a word boundary, it is possible that we inadvertently construct a corrupt spin rate reading by combining incompatible words.

In actual experience, these errors were rare and could be easily filtered out manually, as they resulted in ``one off'' spin rate readings that were wildly off the otherwise steady, slowly changing spin rate value. Moreover, when a suspect spin reading was encountered, we examined the telemetry frame from which the spin word in question was extracted, and often found that the frame contained other corrupt information as well (e.g., wrong format identifier, subcommutator index out of sequence, inconsistent readings in other data words.) As a result, even early in the mission, when spin rate changes were more rapid and frequent, these ``one off'' errors could be easily and unambiguously identified.

\subsection{Pioneer 10 Spin Data}

Spin data for Pioneer 10 came from several sources.

The star sensor on Pioneer 10 failed after Jupiter encounter. Thereafter, the primary spin reference was one of the spacecraft's two sun sensors. The exact time when the switch from the star sensor to the sun sensor took place is not known; spin rate telemetry appears reliable and accurate (jitter free) both before and after Jupiter encounter (11 days worth of telemetry during Jupiter encounter is missing from the telemetry files.)

The Pioneer 10 sun sensor became temporarily inoperative in November 1983 when a CONSCAN maneuver inadvertently reduced the Sun-spacecraft-spin axis angle to $\sim$1/4$^\circ$. The sun sensor was turned back on in July 1985; it was switched off permanently in May 1986, in order to avoid missed sun pulses (and incorrect results after the sensor averaged pulse periods over several spacecraft revolutions.)

When no sun sensor or star sensor was operating, the spacecraft continued to generate ``roll index'' pulses at the rate of the last spin rate measurement.

For Pioneer 10, a total of 98,286 hourly spin readings were extracted from the telemetry. As a result of manual editing, 227 rows were removed from this total, representing approximately 0.33\% of the data collected.

After removing data for the periods when no on-board spin reference was functioning, the remaining data set contains 57,376 rows.

After 1986, spot measurements of the Pioneer 10 spin rate were made using the Imaging Photo-Polarimeter (IPP) instrument \cite{AIAA870502}. In one operating mode, the IPP could be commanded to begin taking an image when a source of sufficient magnitude entered the instrument's field of view. The time from the last roll index pulse until the moment the instrument was triggered could be measured. Taking several such measurements at different times, the difference in time could serve as an indicator of the deviation of the spacecraft's actual spin rate relative to the ``frozen'' rate of the roll index pulse.

These measurements were made periodically until October 1993, when there was no sufficient power left on the spacecraft to continue operating the IPP instrument.

Pioneer 10 spin telemetry data, in units of RPM (revolutions per minute) is presented in the form of a tab-delimited text file, {\tt 23spin\_valid.dat}. This file contains valid data records: i.e., the set of 57,376 spin telemetry records from which bad records and records containing ``frozen'' spin rate readings have been removed. The three columns of the file contain the date ({\tt DOW Mon  D hh:mm:ss yyyy}), the spin rate in RPM, and the contents of telemetry word $C_{431}$ that tells us which sensor was used to obtain the spin rate reading. (From \cite{PC202} we learn that the last two bits of this telemetry word identify the star sensor or either of the two sun sensors as the roll rate reference in use. In actuality, the bits do not seem to be correlated with our knowledge of the spacecraft's state, and may not be trusted.)

A complete set of Pioneer 10 spin data combined from all sources is presented in the Microsoft Excel format data file {\tt 23spin.xls}. The file contains one worksheet with five columns. The first column is the ground received time of the hourly batch from which the spin rate was extracted. The second column (labeled ``Spin (RPM)'') contains the spin rate measurement combined from all three sources. The third, fourth, and fifth columns (labeled ``RollIDX'', ``IPP'', and ``Est'', respectively) contain the information obtained from the on-board roll index pulse reference (i.e., the star sensor or Sun sensors), the IPP instrument, and (after 1993) from ground-based estimates. Note the decreasing accuracy (fewer decimal digits) towards the end of the data set.

\subsection{Pioneer 11 Spin Data}

The star sensor on board Pioneer 11 remained functional throughout the spacecraft's mission, and provided reliable spin telemetry from launch until the last received signal.

The spin rate of Pioneer 11 was highly anomalous. Early in the mission, a thruster malfunction caused Pioneer 11 to spin up to approximately 7.6~rpm. The spin rate of Pioneer 11 continued to increase afterwards. Looking at the telemetry closely, one can observe that the spin rate actually {\it decreased} during quiescent periods between maneuvers (consistently with the behavior of Pioneer 10), but increased suddenly during each maneuver. This is interpreted as a result of a slight misalignment, relative to the spin axis, of one or more maneuvering thrusters. The spin rate increase was significant; at the time when the last spin rate telemetry was received, Pioneer 11 was already spinning at 8.37~rpm.

For Pioneer 11, a total of 90,536 hourly spin readings were collected. During manual editing, 47 records were removed, representing 0.05\% of the total.

Pioneer 11 spin data is presented in the form of a tab-delimited text data file, {\tt 24spin\_edited.dat}. The file contains the 90,489 valid telemetry readings. The columns of the file are the date ({\tt DOW Mon  D hh:mm:ss yyyy}), the spin rate in RPM, and the contents of telemetry word $C_{431}$. (See the previous section for interpreting this word.)

\subsection{Summary}

\begin{table}[ht]
\caption{Data sources used for Pioneer 10/11 spin rate estimation}
\label{tb:spins}
\centering
\begin{tabular}{|c|c|}\hline\hline
{\bf Date}&{\bf Source}\\\hline
\multicolumn{2}{|l|}{\bf Pioneer 10}\\\hline
Launch - December 1973&Star sensor\\
December 1973 - November 1983&Sun sensor\\
November 1983 - July 1985&IPP instrument\\
July 1985 - May 1986&Sun sensor\\
May 1986 - 1993&IPP instrument\\
1993 - EOM&No spin rate measurement\\\hline
\multicolumn{2}{|l|}{\bf Pioneer 10}\\\hline
Launch - EOM&Star sensor\\\hline
\end{tabular}
\end{table}

\begin{figure}[ht]
\hskip -6pt
\begin{minipage}[b]{.5\linewidth}
\centering \psfig{file=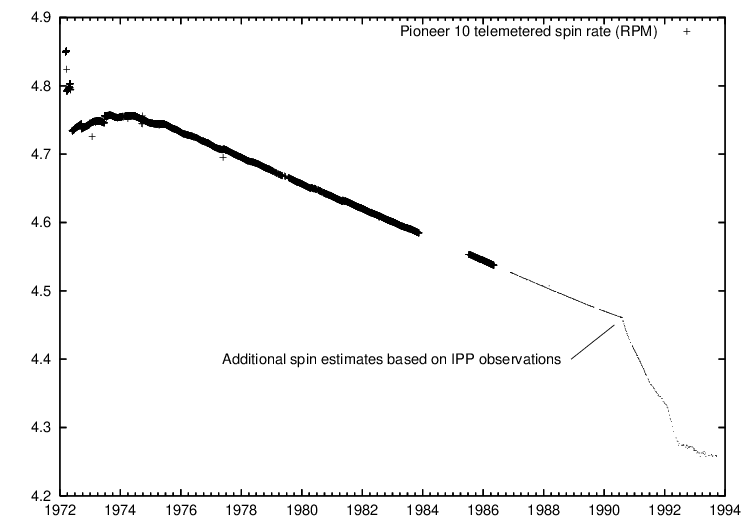, width=\linewidth}
\end{minipage}
\hskip 0.001\linewidth
\begin{minipage}[b]{.5\linewidth}
\centering \psfig{file=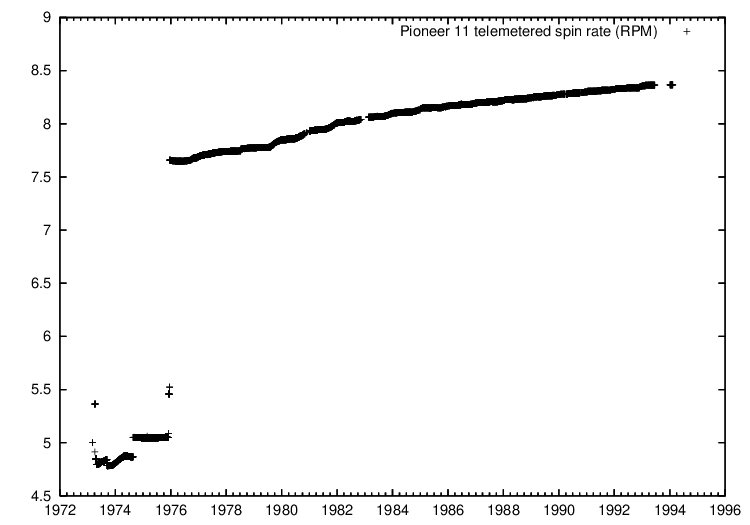, width=\linewidth}
\end{minipage}
\caption{On-board spin rate measurements for Pioneer 10 (left) and Pioneer 11 (right). The sun sensor used on Pioneer 10 for spin determination was temporarily disabled between November 1983 and July 1985, and was turned off in May 1986, resulting in a `frozen' value being telemetered that no longer reflected the actual spin rate of the spacecraft. Continuing spot measurements of the spin rate were made using the Imaging Photo-Polarimeter (IPP) until October 1993. The anomalous increase in Pioneer 11's spin rate early in the mission was due to a failed spin thruster. Continuing increases in the spin rate were due to maneuvers; when the spacecraft was undisturbed, its spin rate slowly decreased (not visible in this plot but evident when one zooms in on a selected date range.)}
\label{fig:spin}
\vskip -5pt
\end{figure}

Table \ref{tb:spins} summarizes the spin data sources that were used for Pioneers 10 and 11. The actual spin readings are show in Figure~\ref{fig:spin}.


\section{Pioneer 10/11 Received Signal Level}

\subsection{Received signal strength telemetry}

The Pioneer telemetry data files contain spacecraft telemetry frames wrapped inside Master Data Records generated by the DSN receiving station. The wrapper record's header and footer contain information about, among other things, the received signal strength and the identifier of the receiving station.

\begin{figure}[ht]
\hskip -6pt
\begin{minipage}[b]{.5\linewidth}
\centering \psfig{file=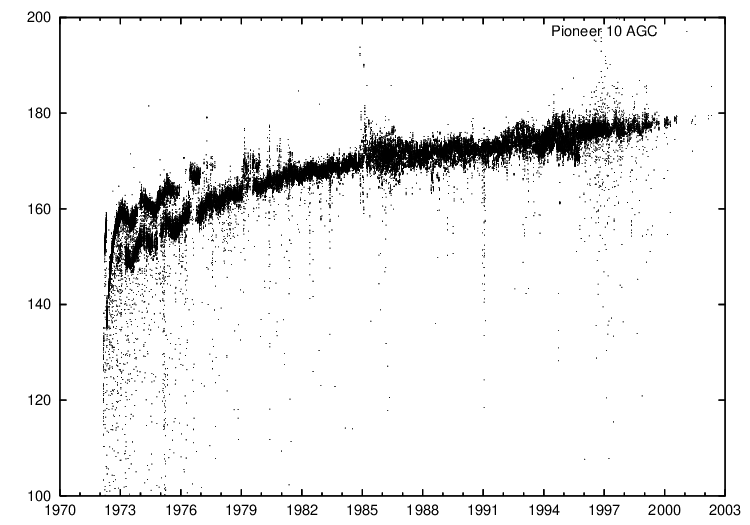, width=\linewidth}
\end{minipage}
\hskip 0.001\linewidth
\begin{minipage}[b]{.5\linewidth}
\centering \psfig{file=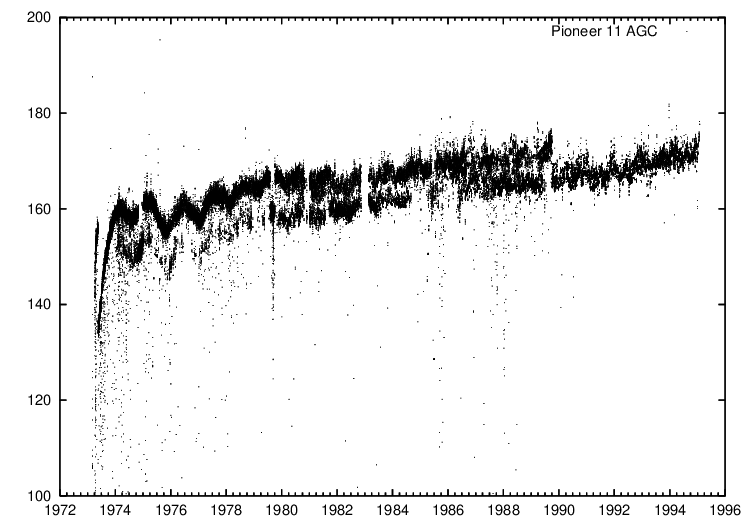, width=\linewidth}
\end{minipage}
\caption{Uncorrected AGC readings from Pioneer telemetry records, averaged hourly.}
\label{fig:AGCraw}
\vskip -5pt
\end{figure}

Specifically, the header contains a fixed-point binary value that is the AGC (Automatic Gain Control) amplification level applied by the DSN station (Figure~\ref{fig:AGCraw}). It is believed that the encoded value is normalized to 0~dBm; in other words, it is the required amplification to bring the received signal to a power level of 1~mW. Therefore, the negative of this number should be interpreted as the actual signal strength in units of dBm.

The data set contains many outliers; we removed all nonsensical data with values outside the range of $1-255$~dB. The result, averaged hourly, is available in the files {\tt 23AGC.dat} and {\tt 24AGC.dat}.

The signal strength depends on several factors, such as:
\begin{itemize}
\item The distance to the spacecraft (space loss)
\item The transmitter signal level on board the spacecraft
\item Pointing error at the spacecraft
\item Pointing error at the DSN
\item Atmospheric loss
\item DSN antenna size
\end{itemize}

Of these, space loss can be calculated using ephemeris data. The transmitter signal level is known from on-board telemetry. The DSN antenna size is also known, although care must be taken to take into account DSN station history, as many stations have undergone upgrades over the years, which in some cases included an enlargement of the antenna dish \cite{UPDOWN}.

The fractional portion of the fixed-point AGC data contains four binary digits, corresponding with a resolution of $1/16$~dB. Translated into distance using the inverse square law, it implies a sensitivity of approximately 0.7\%. Except for the initial few hours of the mission, this value is much larger than the diameter of the Earth, which means that we do not require ephemeris data specific to the DSN station to compensate for space loss; the distance between the spacecraft and the Earth barycenter should be sufficient.

The transmitted signal level is measured on board the spacecraft and telemetered to the ground (telemetry words $C_{231}$ and $C_{214}$.

The DSN antenna diameter is available from historical NASA sources, e.g., \cite{UPDOWN}. The antenna gain is assumed to be proportional to the square of the antenna diameter.

\begin{figure}[ht]
\hskip -6pt
\begin{minipage}[b]{.5\linewidth}
\centering \psfig{file=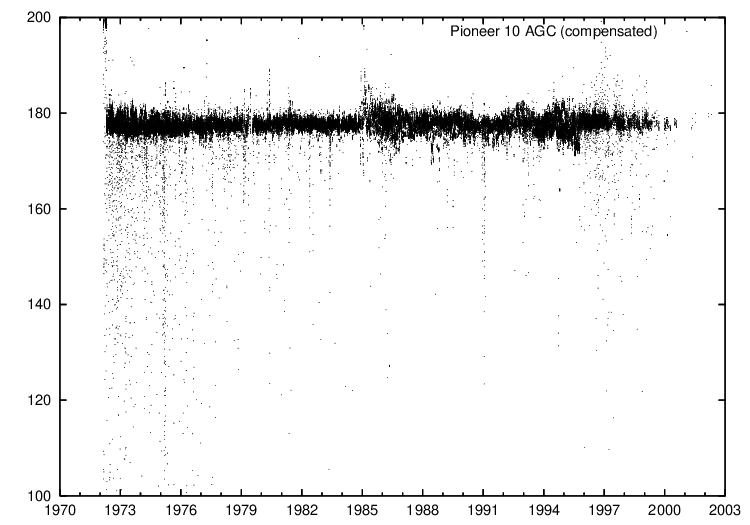, width=\linewidth}
\end{minipage}
\hskip 0.001\linewidth
\begin{minipage}[b]{.5\linewidth}
\centering \psfig{file=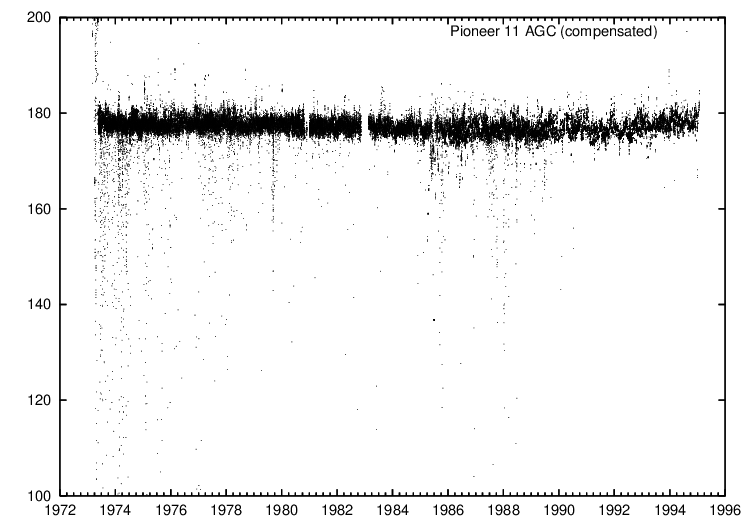, width=\linewidth}
\end{minipage}
\caption{AGC readings from Pioneer telemetry records, averaged hourly, and corrected for space loss, DSN antenna size, and spacecraft transmitter power.}
\label{fig:AGCcomm}
\vskip -5pt
\end{figure}

We normalized the AGC signal (Figure~\ref{fig:AGCcomm}) by compensating for these three factors using the expression
\begin{equation}
P_\mathrm{AGC}'=P_\mathrm{AGC}-20\log_{10} R+P_\mathrm{TWT}+20\log_{10}(D/70),
\end{equation}
where $P_\mathrm{AGC}$ is the measured AGC value, $R$ is the geocentric distance to the spacecraft, and $D$ is the DSN antenna diameter, normalizes the signal relative to a 1~mW transmitter at 1~AU from the Earth, as received by a 70~m antenna. The normalized values are available in the files {\tt 23AGCcomm.dat} and {\tt 24AGCcomm.dat}.

\subsection{Pioneer 10 AGC data}

The transmitter power on board Pioneer 10 was declining over the years, according to on-board telemetry, by as much as 3~dB late in the mission. This decline was clearly visible in the AGC data. After correcting for the transmitter power as well as space loss and DSN antenna size, the resulting curve is essentially flat.

\subsection{Pioneer 11 AGC Data}

The transmitter power on board Pioneer 11 was much more constant than on Pioneer 10; however, there is reason to believe that cumulative pointing errors may have affected communication system performance. This may be visible in the data, which shows a gradual, slight decrease in the received signal strength, a trend that appears to have been slightly reversed about two thirds into the mission.

\subsection{Noise Sources}

The normalized data sets, while showing essentially flat levels as expected, are still quite noisy. What could be causing this noise? First, the DSN antenna gain used in the normalization process was deduced from the DSN antenna nominal size; the actual antenna gain may differ from this value by several dB, causing the normalized signal level to jump as different DSN stations were used for data reception.

Second, there are other sources of noise that were not accounted for. DSN antenna mispointing is a possibility; we may have valid MDRs that were recorded while the DSN was still ``homing in'' on the signal, causing a variation of several dB in the received signal level. Next, there is atmospheric loss; the apparent elevation of the spacecraft above the horizon at the DSN station likely results in a variation in the received signal level, as does the state of the atmosphere/ionosphere at the time of the reception.

It may be possible to compensate for these noise sources using a variety of strategies. Better (historical) information about DSN antenna gain may be available from alternate sources. The data set can be filtered to include data only from specific a DSN station, eliminating the variability due to differences in the normalized antenna gain. Records obtained early or late during a DSN pass could be excluded from the analysis to minimize the effects of atmospheric loss.

Either with or without these improvements, it is hoped that the AGC data set could be used to help solve for spacecraft orientation during orbital analysis, possibly helping with deciding the question of the direction of the Pioneer Anomaly.


\section{English units of measurement}

\begin{table}
\caption{English units of measurement used in Pioneer 10/11 telemetry and project documentation.\label{tb:units}}
\vskip 6pt
\centering
\begin{tabular}{|l|l|c|}\hline
Unit&Description&Conversion\\\hline\hline
$^\circ$F&Degrees Fahrenheit&$x\,^\circ\mathrm{F}=5x/9+255.38\,\mathrm{K}$\\\hline
"&Inches&$1"=0.0254\,\mathrm{m}$\\\hline
ft&Feet&$1\,\mathrm{ft}=0.3048\,\mathrm{m}$\\\hline
sq in&Square inches&$1\,\mathrm{sq\,in}=0.0064516\,\mathrm{m}^2$\\\hline
PSIA&Pounds per sq in absolute&$1\,\mathrm{PSIA}=6894.7573\,\mathrm{Pa}$\\\hline
\end{tabular}
\end{table}

Most Pioneer project documentation, having been produced in the late 1960s and early 1970s, uses English units of measurement. Furthermore, all Pioneer telemetry is calibrated in English units. Rather than converting readings to SI units at all times, sometimes it is more convenient (and less error-prone) to use the original values.

Table~\ref{tb:units} contains a list of Pioneer telemetry units used in this paper and their conversions to SI units of measurement.

\section{Abbreviations}

\newcommand\abbpar[2]{\par\hskip 2em{\bf\vskip 3pt #1} #2}

\begin{longtable}{l l}
{\bf ACS} & Attitude Control Subsystem \\
{\bf A/D} & Analog to digital \\
{\bf AGC} & Automatic gain control \\
{\bf ARC/PA} & Ames Research Center Plasma Analyzer \\
{\bf BPS} & Bits per second \\
{\bf CDU} & Command distribution unit \\
{\bf CEA} & Control electronics assembly \\
{\bf CIT/IR} & California Institute of Technology Infrared Radiometer \\
{\bf CTRF} & Central transformer-rectifier-filter \\
{\bf dBm} & Decibel over miliwatt \\
{\bf DDA} & Digital data acquisition \\
{\bf DDT} & Data-dependent type code \\
{\bf DDU} & Digital decoder unit \\
{\bf DHS} & Data Handling Subsystem \\
{\bf DSL} & Duration and steering logic \\
{\bf DSN} & Deep Space Network \\
{\bf DSS} & DSN Station \\
{\bf DSU} & Data storage unit \\
{\bf DTU} & Digital telemetry unit \\
{\bf DQI} & Data quality indicator \\
{\bf GDD} & Gross data descriptor \\
{\bf GE/AMD} & General Electric Asteroid/Meteoroid Detector \\
{\bf GSFC/CRT} & Goddard Spacecraft Flight Center Cosmic Ray Telescope \\
{\bf HGA} & High-gain antenna \\
{\bf HSD} & High speed data \\
{\bf JPL/HVM} & Jet Propulsion Laboratory Helium vector magnetometer \\
{\bf LaRC/MD} & Langley Research Center Meteoroid Detector \\
{\bf LGA} & Low-gain antenna \\
{\bf LSB} & Least significant bit(s) \\
{\bf MDR} & Master data record \\
{\bf MGA} & Medium gain antenna \\
{\bf MSB} & Most significant bit(s) \\
{\bf PSA} & Propellant supply assembly \\
{\bf PSE} & Program storage and execution \\
{\bf PSIA} & Pounds per square inch absolute \\
{\bf RPM} & Revolutions per minute \\
{\bf RTG} & Radioisotope thermoelectric generator \\
{\bf S/C} & Spacecraft \\
{\bf SCF} & SCID correction flag \\
{\bf SCID} & Subcommutator identifier \\
{\bf SCT} & Spin control thruster \\
{\bf SNR} & Signal-to-Noise Ratio \\
{\bf SPSG} & Spin period sector generator \\
{\bf SRA} & Stellar reference assembly \\
{\bf SSA} & Sun sensor assembly \\
{\bf TRF} & Transformer, rectifier, and filter \\
{\bf TWT(A)} & Traveling wave tube (amplifier) \\
{\bf UA/IPP} & University of Arizona Imaging Photo-Polarimeter \\
{\bf UC/CPI} & University of Chicago Charged Particle Instrument \\
{\bf UCSD/TRD} & University of California at San Diego Trapped Radiation Detector \\
{\bf UDT} & User data type code \\
{\bf UI/GTT} & University of Iowa Geiger Tube Telescope \\
{\bf USC/UV} & University of Southern California Ultra-Violet photometer \\
{\bf VCO} & Voltage controlled oscillator \\
{\bf VPT} & Velocity and precession thruster
\end{longtable}

\bibliography{refs}
\bibliographystyle{unsrt}

\end{document}